%% file: main.tex
\documentclass[12pt, 3p, numbers, authoryear]{elsarticle}
\usepackage{lipsum} 	%style package
\usepackage{graphicx}	%images package
\usepackage{float}		%float positions

\usepackage{booktabs}
\usepackage[normalem]{ulem}
\useunder{\uline}{\ul}{}

\usepackage{longtable,booktabs}
\usepackage{multirow}
\usepackage[table,xcdraw]{xcolor}
\usepackage{mathtools}
\usepackage{amssymb}
\usepackage{amsmath}
\usepackage{caption}
\usepackage{subcaption}
\usepackage[position=top]{subfig}
\usepackage{tikz}

\usepackage{framed} % Framing content
\usepackage{nomencl} % Nomenclature package
\makenomenclature
\usepackage{etoolbox}
\renewcommand\nomgroup[1]{%
  \item[\bfseries
  \ifstrequal{#1}{L}{List of Symbols}{%
  \ifstrequal{#1}{P}{Superscripts}{%
  \ifstrequal{#1}{B}{Subscripts}{}}}%
]}
\makeatletter
\newcommand{\vast}{\bBigg@{4}}
\newcommand{\Vast}{\bBigg@{7}}
\makeatother

\usepackage{fancyhdr}
\begin{document}
%\maketitle

\begin{frontmatter}

\title{Flow-induced breakup of drops and bubbles}

\author[label1]{Suhas Jain S\fnref{label2}}
\address[label1]{Institute of Fluid Dynamics, Helmholtz-Zentrum Dresden-Rossendorf, Germany}

\ead{sjsuresh@stanford.edu}
\fntext[label2]{Presently a graduate student at Center for Turbulence Research, Stanford University, USA}

\begin{abstract}
\input{abstract.tex}
\end{abstract}

\begin{keyword}
%% keywords here, in the form: keyword \sep keyword
drop \sep bubble \sep viscous \& inertial breakup \sep breakup time \sep critical diameter \sep maximum deformation
%% MSC codes here, in the form: \MSC code \sep code
%% or \MSC[2008] code \sep code (2000 is the default)
\end{keyword}

\end{frontmatter}

%Front matter
%\pagenumbering{roman}
%\section*{Summary}
%\addcontentsline{toc}{section}{\numberline{}Summary}
%\cleardoublepage

\newpage

%Table of contents
%\tableofcontents
%\thispagestyle{empty}
%\cleardoublepage

%List of figures
%\listoffigures
%\addcontentsline{toc}{section}{\numberline{}List of Figures}
%\cleardoublepage

%List of tables
%\listoftables
%\addcontentsline{toc}{section}{\numberline{}List of Tables}
%\cleardoublepage

%Body of the article
\pagenumbering{arabic}
\setcounter{page}{1}

\input{introduction.tex}

\input{viscous.tex}

\input{inertial.tex}

\input{turbulent.tex}

\input{references.tex}

\end{document}

%% file: abstract.tex
Breakup of drop/bubble can be viewed as a result of fundamental force balance when the disruptive force is greater than the restorative force. A disruptive force acting on the drop/bubble tries to deform it, whereas a restorative force refrains it from deforming. Studying breakup and coalescence phenomenon is utmost important since it governs the amount of interfacial area and hence the exchange of heat, mass and momentum across the interface. It also helps in the development of better closure relations for modeling large scale systems.

In this paper, abundant literature consisting of theoretical, experimental and numerical works up to date is reviewed. Broadly, breakup is classified into viscous, inertial and complex turbulent breakup. Physics involved and non-dimensional numbers governing the drop and bubble breakup in various flow configurations are discussed. Characteristic parameters of the breakup such as critical diameter ($d_{max}$), maximum deformation, breakup time ($t_b$), wavelength of disturbance ($\lambda$), impurities in the flow, initial shape of the drop/bubble, history of the flow and critical values of non-dimensional numbers are examined and the important parameters are listed for ready-to-use in modeling approaches. Finally, scope for future work in number of areas is identified.

%% file: introduction.tex
\section{Introduction}\label{introduction}

Flow of a continuous liquid phase containing bubbles or immiscible drops
occurs frequently in a large variety of natural phenomena and technical
processes. Examples comprise emulsification, extraction, absorption,
distillation and boiling which are relevant to food processing, chemical
engineering and energy production among others. An understanding of such
flows therefore is of high interest since it can lead to predictive
models for CFD-simulation that would be useful for scale-up and
optimization of the mentioned processes. Progress in this direction is
difficult because a wide range of scales is involved with the size of
the domain occupied by the two-phase mixture at the large end and the
size of the individual drops or bubbles at the small end. Moreover a
spectrum of drop or bubble sizes is typically present.

In dispersed multiphase flows, knowledge of the distribution of drop or
bubble sizes is a central issue because this distribution determines the
interfacial area which in turn determines the interfacial mass,
momentum, and energy exchange. Concerning multiphase CFD-simulation
these exchange processes have to be modeled within the Eulerian
two-fluid framework of interpenetrating continua that is required to
facilitate computation on large domains for industrial applications. The
evolution of the drop or bubble size distribution in turn is governed by
the dynamics of coalescence and breakup processes. Many attempts have
been made to include closure models for these processes in the
Euler-Euler description of dispersed two-phase flows (Liao and Lucas
2009, 2010, Solsvik et al. 2013). However, no agreement on a suitable
one has been reached and the predictive capability is far from
satisfactory. In part this state of affairs is due to the fact that both
processes compete and only the difference of their rates is observable
as a change in the size distribution.

In the present work we take a step back and focus on the available
mechanistic understanding of these processes that forms the basis of any
modeling work. Of the two, understanding of coalescence appears more
advanced to date than understanding of breakup. In addition, coalescence
can be suppressed by suitable additives if desired, while no such
possibility is known for breakup. Therefore, a viable strategy for
Euler-Euler closure modeling is to first consider breakup individually
and validate its description for cases where coalescence is suppressed.
In a second step the description of coalescence can then be added and
validated for cases where both processes occur by keeping the previously
developed breakup models fixed. Accordingly, the present work will be
limited to the first step and consider only breakup.

Research on deformation and breakup of drops or bubbles has progressed
significantly starting from the works of Taylor (1932), but a complete
understanding is still not available. Even today new mechanisms causing
breakup are being observed both numerically and experimentally for which
adequate models are yet to be developed and validated.

A major distinction between the acting mechanisms may be made based on
the Reynolds number of the flow. At low Reynolds number, breakup is due
to the viscous shear stresses in the external fluid and the breakup
depends on the Capillary number, initial shape of the particle
undergoing breakup, viscosity ratio and in some cases on the
concentration of the impurities in the system. At higher Reynolds
number, breakup is due to the inertial forces acting on the particle and
the breakup depends on the Weber number, Ohnesorge number, viscosity
ratio and on the shape of the particle undergoing breakup. In the case
of turbulent flows, the occurrence of breakup also depends on the
complete history of the flow.

Breakup in simple flows is rather well understood. The focus of most of
the research in this area has been on the prediction of the critical
conditions beyond which no steady drop or bubble shape exists. Breakup
criteria are known from the experimental results/ theoretical/ numerical
models and these serve as the starting point for the further research.
Though this method was considered as reasonably successful by various
authors, it has failed in describing accurately the critical conditions
for breakup.

The current state of understanding of breakup in complex flows in
contrast is still meager. For example, the critical Weber number in
Kolmogorov-Hinze theory (obtained by balancing the external force with
the surface tension force) alone fails to explain the Resonance breakup.
Hence further study is required to determine the variables affecting the
breakup process in various conditions. The breakup dynamics of the
particles in turbulent flow with buoyancy is found to be different from
that without buoyancy as observed by Ravelet, Colin and Risso (2011).
More detailed studies are yet to be done.

The objective of the present work is to provide a comprehensive review
of the literature on the breakup of drops and bubbles. It aims at
systematically categorizing and describing all the mechanisms involved
in a complete understanding of breakup phenomena, which is crucial for
the understanding of dispersed multiphase flows and the development of
closure relations for two-fluid models. In this way also a list of the
relevant parameters influencing the breakup is obtained. In particular,
new findings on breakup in turbulent flows are emphasized and needs for
future work in this area are identified. Inevitably there is some
overlap with the previous reviews of Stone (1994), Gelfand (1996), Risso
(2000), and Ashgriz (2011) to make this article self-contained. The
presentation is limited to simple Newtonian fluids. Topics such as
breakup up in the presence of surfactants or in electric and magnetic
fields or non-Newtonian effects are not considered here. A review on
surfactant effects has been given by Stone \& Leal (1990), the presence
of electric fields has been considered e.g. by Ha \& Yang (2000), and
non-Newtonian effects are discussed in the book by Chabra (1990, Chap.
6.8).

Drops and bubbles share many features. In fact, they are distinguished
only by values of the density- and viscosity-ratio relative to the
continuous liquid being small (for bubbles) or not (for drops) which
really means a gradual variation. For air bubbles in water for example
the ratio of the dynamic viscosities of air to water is about 0.02. As
will be seen this is not negligible in some circumstances. The term
``particle'' will therefore be used further on to refer to either a drop
or bubble. As much as possible results are expressed in terms of
dimensionless numbers.

Frequently appearing ones are the Reynolds number, Weber number,
Capillary number, Ohnesorge number, Bond number (E\"{o}tv\"{o}s number),
Archimedes number, Galilei number, viscosity-ratio, where the viscosity
ratio is defined as the ratio of the dynamic viscosity of the internal
fluid to that of the external fluid,
\\
\begin{equation}
    \lambda=\frac{\mu_d}{\mu_c}
\end{equation}
\\
Reynolds number is defined as the ratio of momentum force to viscous force, 
\\
\begin{equation}
    Re=\frac{\rho_c U d}{\mu_c}
\end{equation}
\\
Weber number is defined as the ratio of inertial force to surface tension force,
\\
\begin{equation}
    We=\frac{\rho_c U^2 d}{\sigma}
\end{equation}
\\
Ohnesorge number is defined as the ratio of viscous force to the geometric mean of inertial and surface tension force,
\\
\begin{equation}
    Oh=\frac{\mu_d}{\sqrt{\rho_d  \sigma d}} = We^\frac{1}{2} Re^{-1}
\end{equation}
\\
Capillary number is defined as the ratio of the viscous force to that of surface tension force acting across an interface,
\\
\begin{equation}
    Ca=\frac{\mu_c U}{\sigma} = We Re^{-1}       
\end{equation}
\\
Bond number (E\"{o}tv\"{o}s number) is defined as the ratio of the body force to that of the surface tension force acting across an interface,
\\
\begin{equation}
    Eo=Bo=\frac{\Delta \rho g d^2}{\sigma}
\end{equation}
\\
Archimedes number (Galilei number) is defined as the ratio of the buoyancy force to the viscous force,
\\
\begin{equation}
    Ga=Ar=\frac{\rho_c g^\frac{1}{2} d^\frac{3}{2}}{\mu_c}(1-\frac{\rho_c}{\rho_c})
\end{equation}

%% file: viscous.tex
\section{Viscous force driven
breakup}\label{viscous-force-driven-breakup}

At low $Re$ values, inertial forces on the particle can be neglected and the particle deformation and breakup are primarily due to the viscous shear stresses. This situation generally applies to the breakup of small particles in highly viscous liquids. Acrivos (1983), Rallison (1984), Stone (1994), and Risso (2000) have extensively reviewed this area.

When inertia is negligible, flow is governed by the Stokes'
equations. Dimensional analysis of these governing equations reveals that the flow is governed majorly by the $\lambda$ and $Ca$. Close observation at results of the theoretical and experimental studies has also revealed that the history of the external force and the initial shape of the particle also play a major role. Further, breakup of this kind can be classified as steady and unsteady based on the characteristics of external flow:

\input{steady_viscous.tex}

\input{unsteady_viscous.tex}

%% file: steady_viscous.tex
\subsection{Steady flows}\label{steady-flows}

\begin{figure}[H]
\centering
\includegraphics[width=5in]{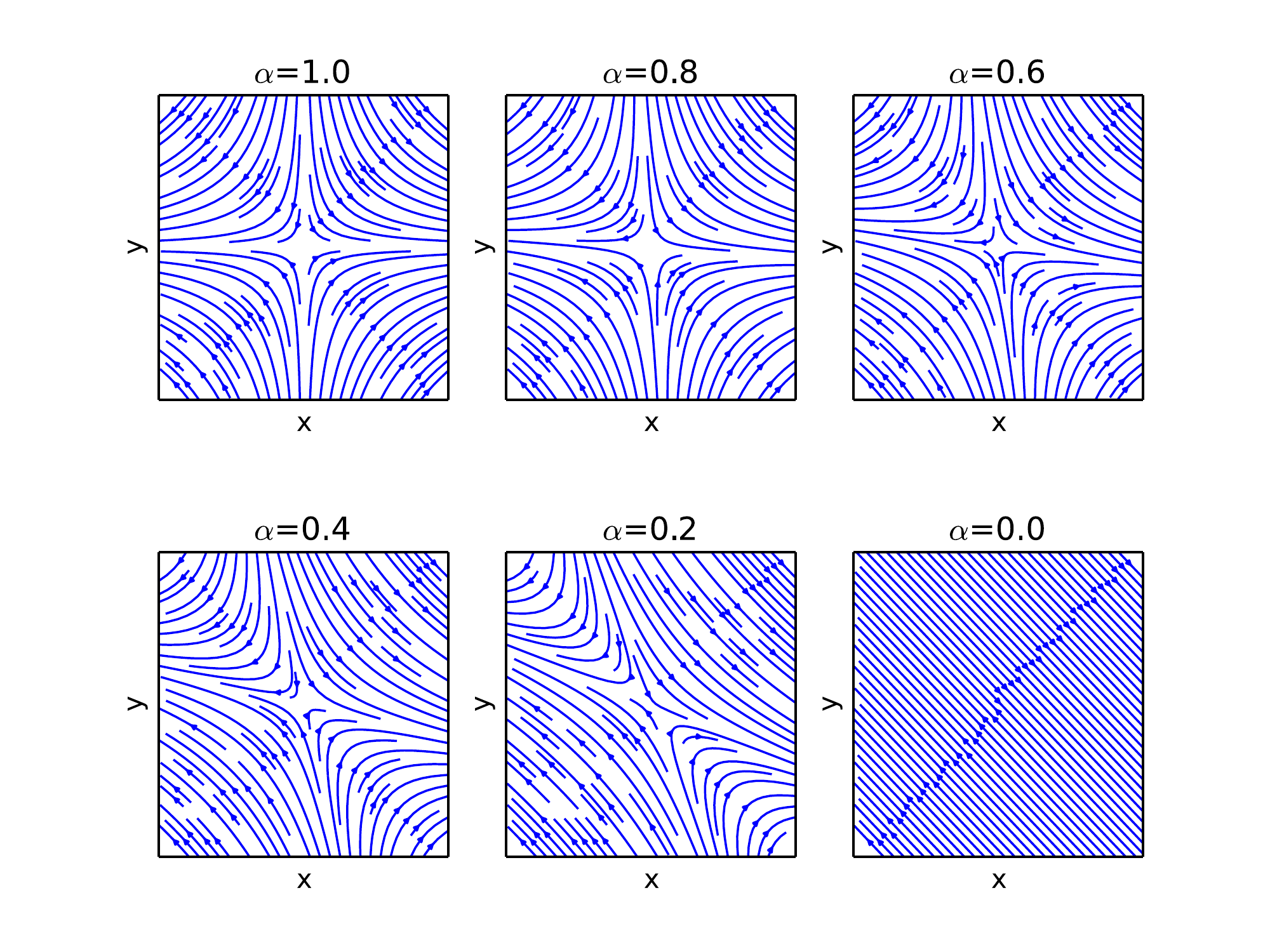}
\centering
\caption{Streamlines of flow field for
\(\mathbf{\alpha}\mathbf{\geq 0}\).}
\label{fig:streamlines}
\end{figure}

This section deals with the breakup in quasi-steady flows. Plane flows of the form
\\
\begin{equation}
    \nabla U = \frac{G}{2} 
    \begin{bmatrix} 
    1+\alpha & 1-\alpha & 0 \\ 
    -1+\alpha & -1-\alpha & 0 \\ 
    0 & 0 & 0  
    \end{bmatrix}
    \label{equ:flow}
\end{equation}
\\
are considered, where \(G\) is the strength of the external flow and
\(\alpha\) is a flow type parameter such that \(\alpha\) = +1 results in  a pure straining flow, \(\alpha\) = 0 a simple shear flow and
\(\alpha\) = -1 a purely rotational flow. Examples of streamline
patterns for \(\alpha \geq 0\) are shown in Figure \ref{fig:streamlines}. Values of
\(\alpha < 0\) have not been considered in this review.

\begin{figure}[H]
\centering
\includegraphics[width=4in]{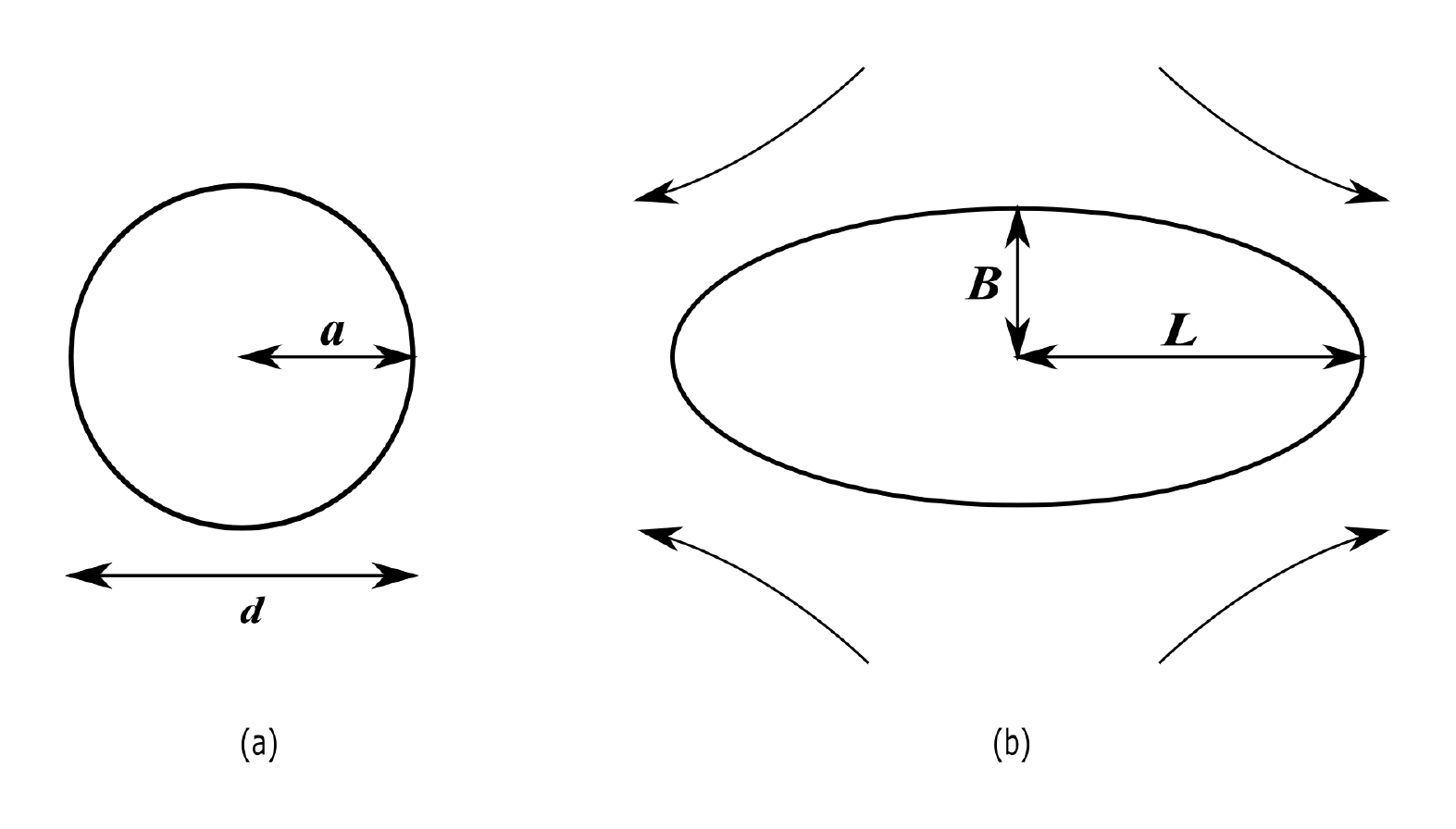}
\centering
\caption{Particle morphology. (a) Initial spherical particle.
(b)Particle shape under the influence of external flow field.}
\label{fig:morphology}
\end{figure}

If exposed to a flow of the type above an initially spherical particle assumes a prolate ellipsoidal shape as sketched in Figure \ref{fig:morphology}. The measure
of deformation $De=(L-B)/(L+B)$ is frequently used for small deformations. At large deformations, it tends to a limiting value of 1 by definition so it ceases to be useful. In this case $\frac{L}{a}=(\frac{1+De}{1-De})^{\frac{2}{3}}$ (assuming a prolate ellipsoidal shape) has to be used, which
has no restrictions on its applicability.

Viscosity ratio has a marked effect on the morphology of the particles. Hence based on this, study is further divided into high, medium and low viscosity ratio flows.

\subsubsection{High viscosity ratio
flows}\label{high-viscosity-ratio-flows}

At high viscosity ratios, \(\lambda\sim 1\) or larger, the disruptive
viscous forces tend to breakup the particles at relatively low shear
rates or at moderate deformations.

Very first studies on this were done by Taylor (1932, 1934). He considered an almost spherical particle in a simple shear flow (\(\alpha\) = 0) and calculated the flow field around it by balancing the largest value of the pressure difference occurring on the particle surface with the average stresses due to surface tension. He obtained an estimate for the maximum stable initial radius of
the particle, given by
\\
\begin{equation}
    a=\frac{2\sigma(\mu_d + \mu_c)}{G(1+\alpha)\mu_c(\frac{19}{4}\mu_d+4\mu_c)}
\end{equation}
\\
which can be rewritten in terms of critical capillary number as
\\
\begin{equation}
    Ca_{crit}=\frac{2(\lambda+1)}{(1+\alpha)(\frac{19}{4}\lambda + 1)}
\end{equation}
\\
Taylor (1934) also performed experiments using a parallel band apparatus on the deformation of a particle of one fluid in another with controlled interfacial tension, viscosities, and rate of deformation of the outer fluid and showed that the experiments matched well with his theory (Taylor 1932) at low shear rates.

Further, Cox (1969) worked on theoretically determining the shape of the particle in simple shear and pure straining flows by assuming the
particle deformation to be small and of order \(\epsilon << 1\) and expanding the velocity field in terms of this parameter. But their model failed to indicate the possibility of breakup. Torza et al. (1972) conducted experiments on
neutrally buoyant liquid particles suspended in viscous liquids
undergoing simple shear flow and validated
the model developed by Cox (1969).

Cox (1969)'s theoretical work was further extended by Frankel and
Acrivos (1970), Barthes-Biesel (1972) and Barthes-Biesel and Acrivos
(1973). They used linear stability theory to predict the onset of
breakup of particles freely suspended in a simple shear flow and found
that the results were in good agreement with the experimental works of
Taylor (1934), Rumscheidt and Mason (1961) and Grace (1971).

\begin{figure}[H]
\centering
\includegraphics[width=5in]{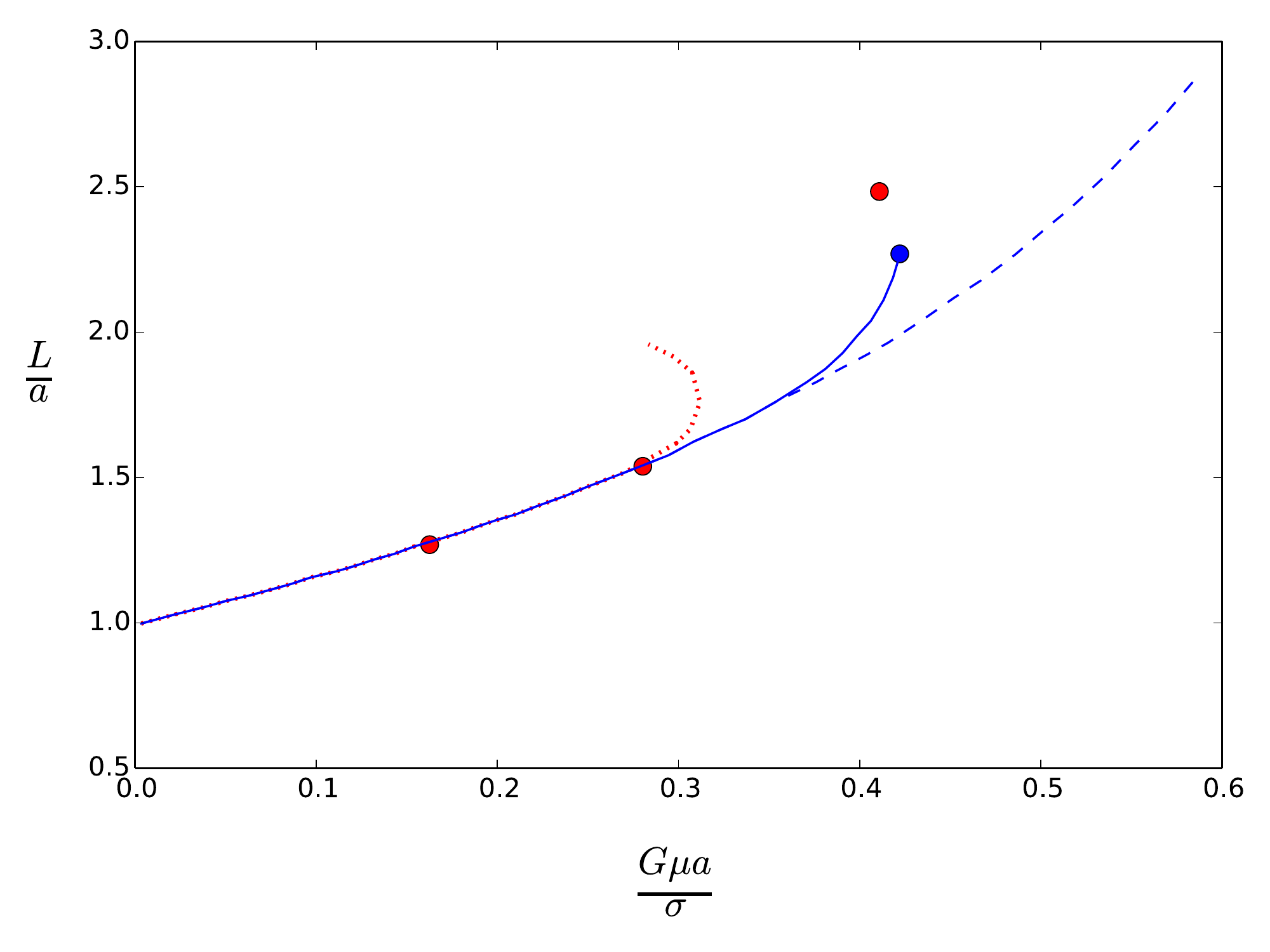}
\centering
\caption{Elongation ratio \(\frac{L}{a}\) of a particle in simple shear
flow when \(\lambda=1\) {[}Redrawn from Figure 4 in
Acrivos (1983){]} . Solid line: numerical results obtained by Rallison
(1981). Filled blue circle: the numerically determined point of breakup.
Filled red circles: experimental measurements by Rumscheidt and Mason
(1961). Dashed line: theory by Taylor (1932). Dotted line: theory by
Barthes-Biesel and Acrivos (1973).}
\label{fig:high_viscosity_ratio}
\end{figure}

Figure \ref{fig:high_viscosity_ratio} summarizes the works on high viscosity ratio flows,

\begin{itemize}
\item
  Taylor's theory is accurate in determining the deformation up to the
  point of breakup, though it doesn't predict critical shear rate for
  breakup.
\item
  Theory by Barthes-Biesel and Acrivos (1973) over predicts the
  deformation but describes the critical shear rate at an acceptable
  accuracy for $\lambda \geq 0.05$ (provides a good qualitative picture but not in good
  quantitative agreement).
\item
  Breakup was obtained by a non-existence of a steady solution for the
  system of equations.
\item
  There is excellent agreement between the calculated deformations and
  those found experimentally, up to the point of breakup.
\end{itemize}

\subsubsection{Low viscosity ratio
flows}\label{low-viscosity-ratio-flows}

On the other hand, for low viscosity ratios, the particles are highly
elongated and slender, i.e. $L$ \textgreater{}\textgreater{} $B$. Taylor
(1964) was the first to calculate the particle deformation in the
slender body limit. He calculated the effect of presence of a particle
on the pure straining flow by a distribution of singularities along the
particle axis and calculated the type, strength and position of these
singularities together with the position of the particle axis that
matched the boundary conditions. Buckmaster (1972, 1973) refined this
theory considering inviscid particles in axisymmetric pure straining
flow of a highly viscous fluid and further Acrivos and Lo (1978)
extended it for low but finite viscosity particles. Hinch and
Acrivos (1980) solved for simple shear flows and Khakhar and Ottino
(1986) have also contributed to this theory.

\begin{figure}[H]
\centering
\includegraphics[width=5in]{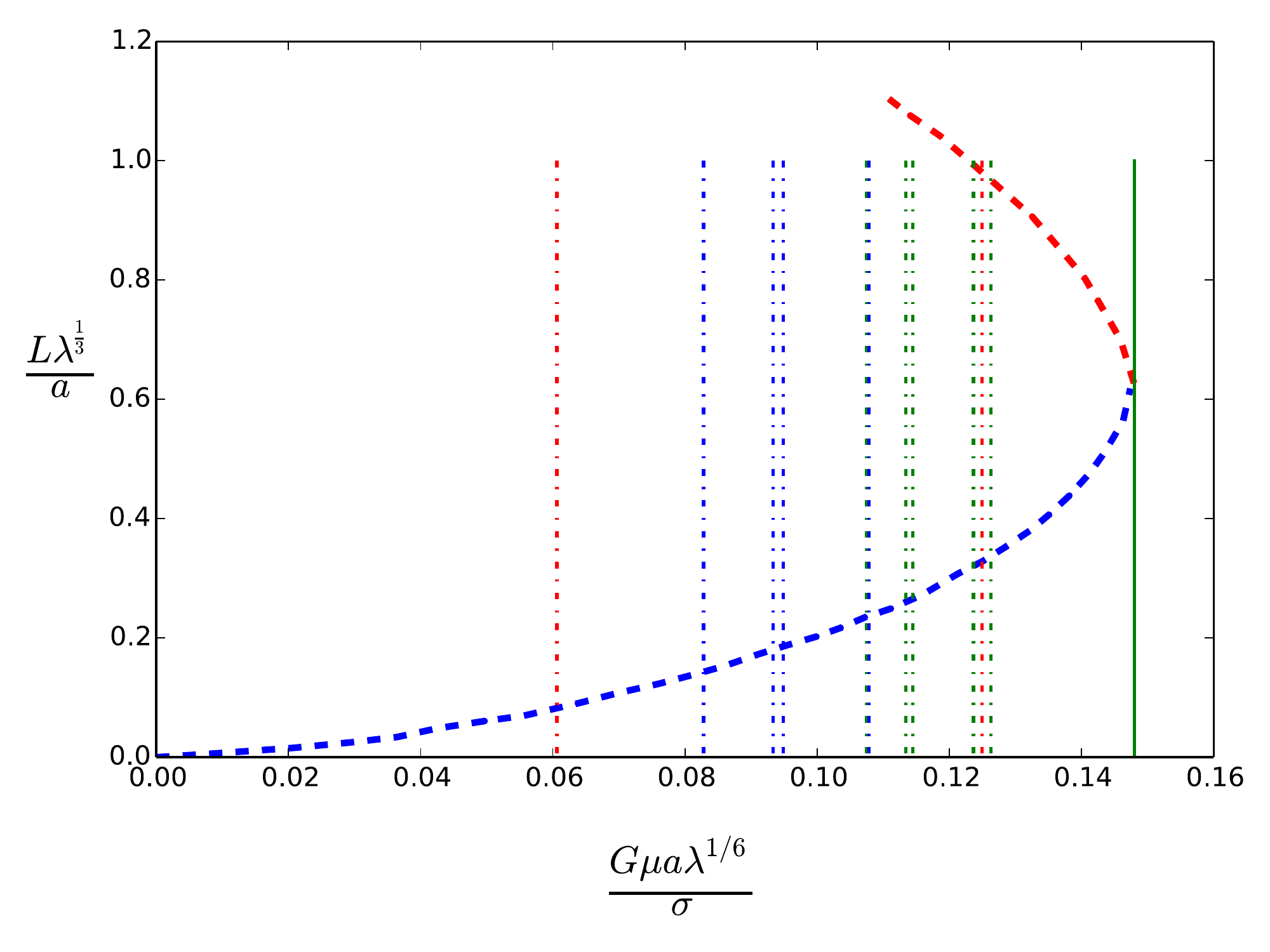}
\centering
\caption{Blue and red dased line: deformation and breakup of a particle
of very low viscosity (\(\lambda\ll 1)\) in pure
straining flow, Solid green line: critical value of \(G\) for
breakup, Blue dash-dotted line: experimental critical value of \(G\) for
breakup at $\mu_c=45.5P$ by Grace (1971), Red dash-dotted line: experimental
critical value of \(G\) for breakup at $\mu_c=502.5P$ by Grace (1971), Green dash-dotted
line: experimental critical value of \(G\) for breakup at $\mu_c=502.5P$ by Yu
(1974) {[}Redrawn from Figure 6 in Acrivos (1983) and Figure 4 in Hinch
and Acrivos (1979){]}.}
\label{fig:low_viscosity_ratio}
\end{figure}

Slender body theory only applies to $\lambda$ \textless{}\textless{} 1 at the
main central region of the particle and not near the pointed ends of the
particle. The deformation relation is given by
\\
\begin{equation}
    10\Big(\frac{G \mu a \lambda^{1/6}}{\sigma}\Big)^2 = \frac{L \lambda ^\frac{1}{3}}{a}/\bigg(1+\frac{4}{5}\Big(\frac{L\lambda^{1/3}}{a}\Big)^3\bigg)^2
\end{equation}
\\
and is shown in Figure \ref{fig:low_viscosity_ratio}.

A more physical explanation is that an increase in the shear rate \(G\)
leads to a decrease in the local cross-sectional radius of the particle,
since the capillary forces must balance the external shear stress; this,
in turn, leads to an increase in \(L\) and corresponding decrease in the
pressure within the particle at its center, which requires a further
decrease in the radius. Clearly, this cannot continue indefinitely and
for large enough values of \(G\) the particle breaks.

Hinch and Acrivos (1979, 1980) extended this analysis to pure straining
flows and to simple shear flows. In the former case, they found that,
although the cross-section of the particle becomes significantly
distorted from the circular shape, the deformation curve remained
essentially similar as in axisymmetric flows and breakup is predicted to
occur when
\(\frac{G\mu a \lambda^{1/6}}{\sigma}\)
reaches a value equal to 0.145, rather than 0.148, as in axisymmetric
flow and that there is good agreement between this theoretical criterion
for breakup and the experimental observations as shown in Figure \ref{fig:low_viscosity_ratio}. In
the latter case of simple shear flow, the situation was much more complex
and only qualitative agreement with the experiment was obtained.

\subsubsection{Medium viscosity ratio
flows}\label{intermediate-viscosity-ratio-flows}

\begin{figure}[H]
\centering
\includegraphics[width=5in]{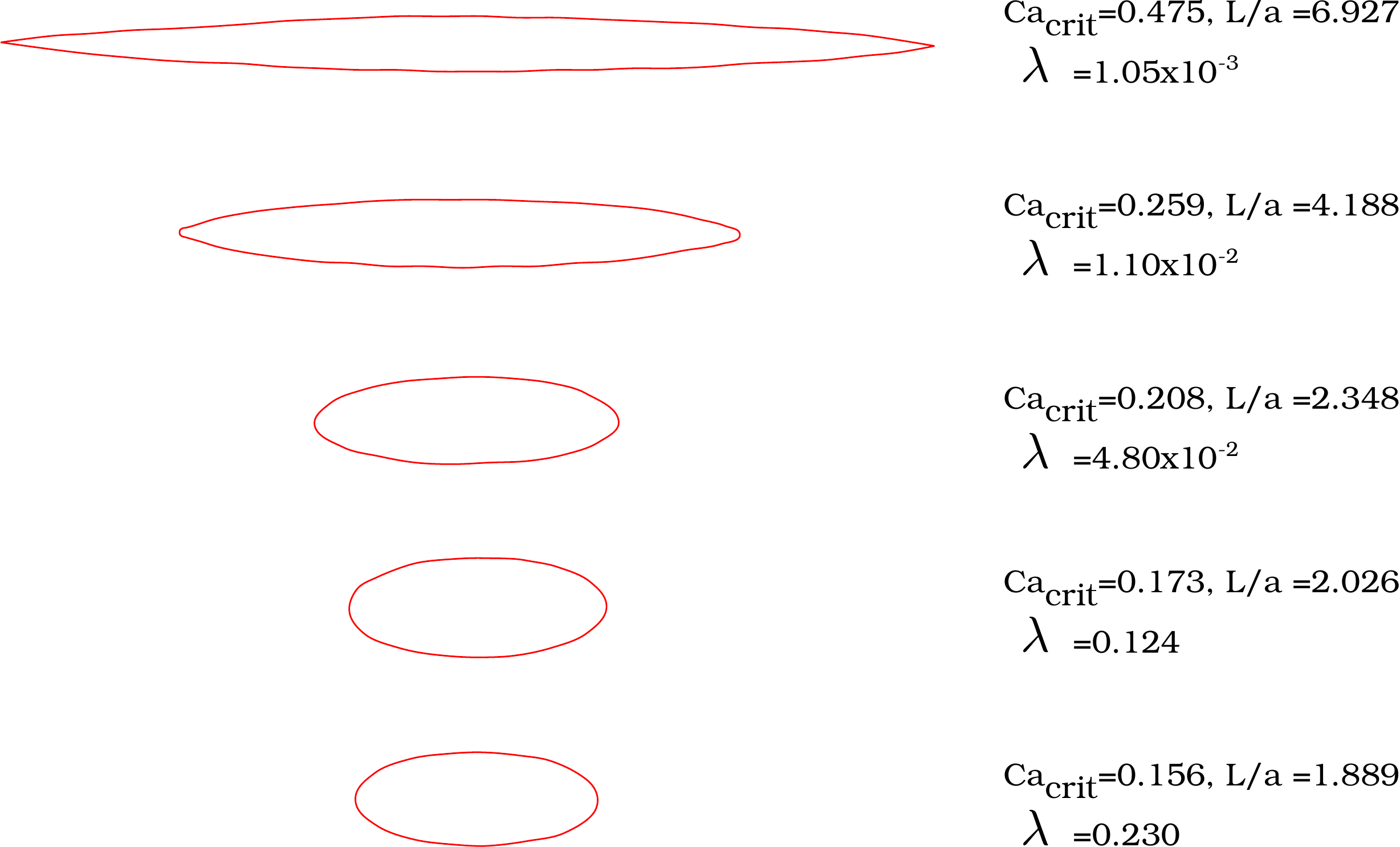}
\centering
\caption{Most extended stable shapes for each
viscosity ratio investigated at \(\mathbf{\alpha = 1.}\) {[}Outlines extracted from experimental photographs of Bentley and Leal (1986){]}.}
\label{fig:extended_shapes}
\end{figure}

\begin{figure}[H]
\centering
\includegraphics[width=5in]{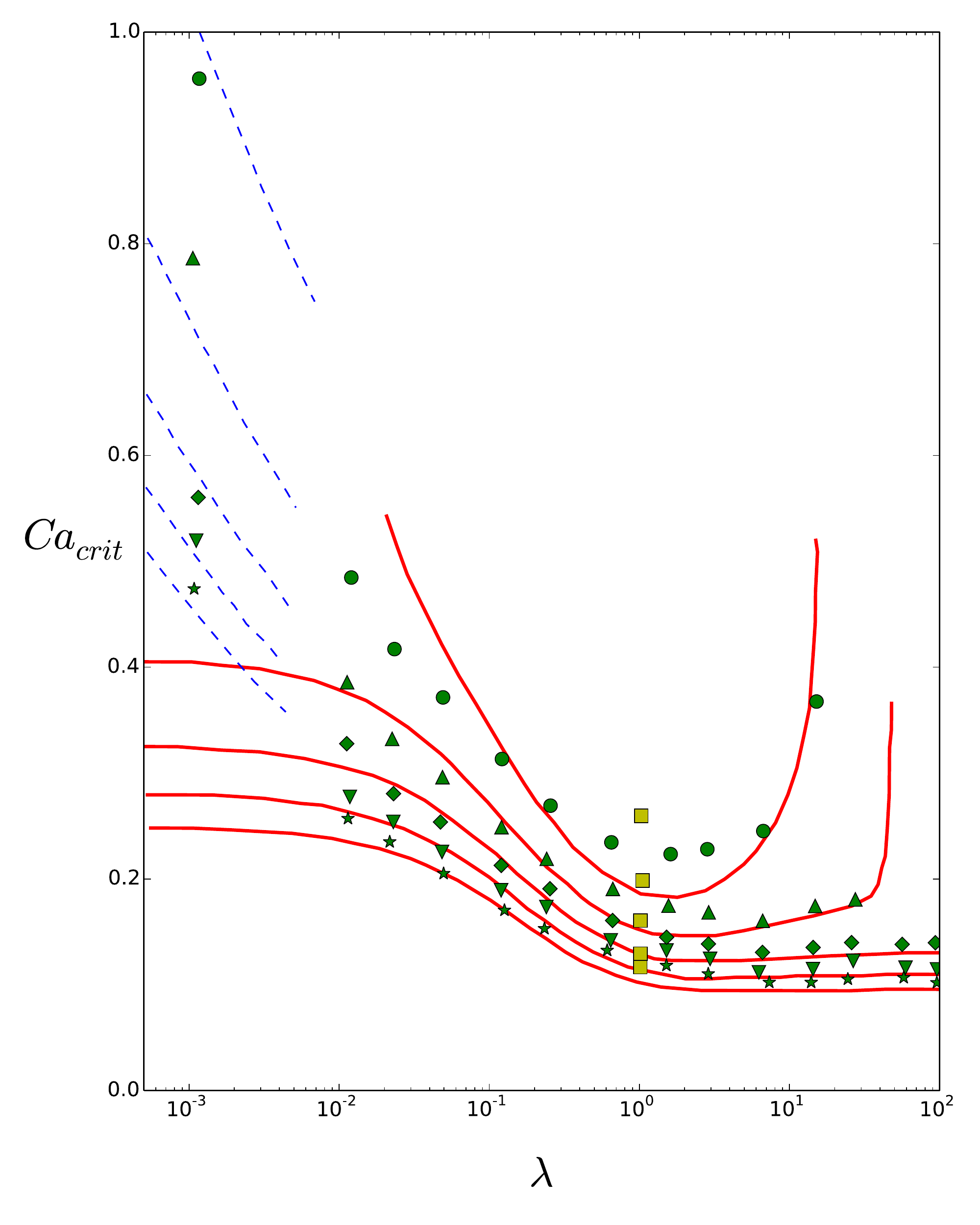}
\centering
\caption{Critical capillary number vs viscosity ratio for different
values of \(\mathbf{\alpha}\). Values of
\(\alpha\ (0.2,0.4,0.6,0.8\ and\ 1.0)\) increases from top to
bottom. Solid line: small-deformation theory, Dashed line:
large-deformation theory, Green circles: \(\alpha = 0.2\),
Green upward triangles: \(\alpha = 0.4\), Green diamonds:
\(\alpha = 0.6\), Green downward triangles:
\(\alpha = 0.8\), Green stars: \(\alpha = 1.0\),
Yellow squares: results obtained by Rallison (1981) {[}Redrawn from the
Figure 28 in Bentley and Leal (1986){]}.}
\label{fig:critical_capillary}
\end{figure}

\begin{figure}[H]
\centering
\includegraphics[width=5in]{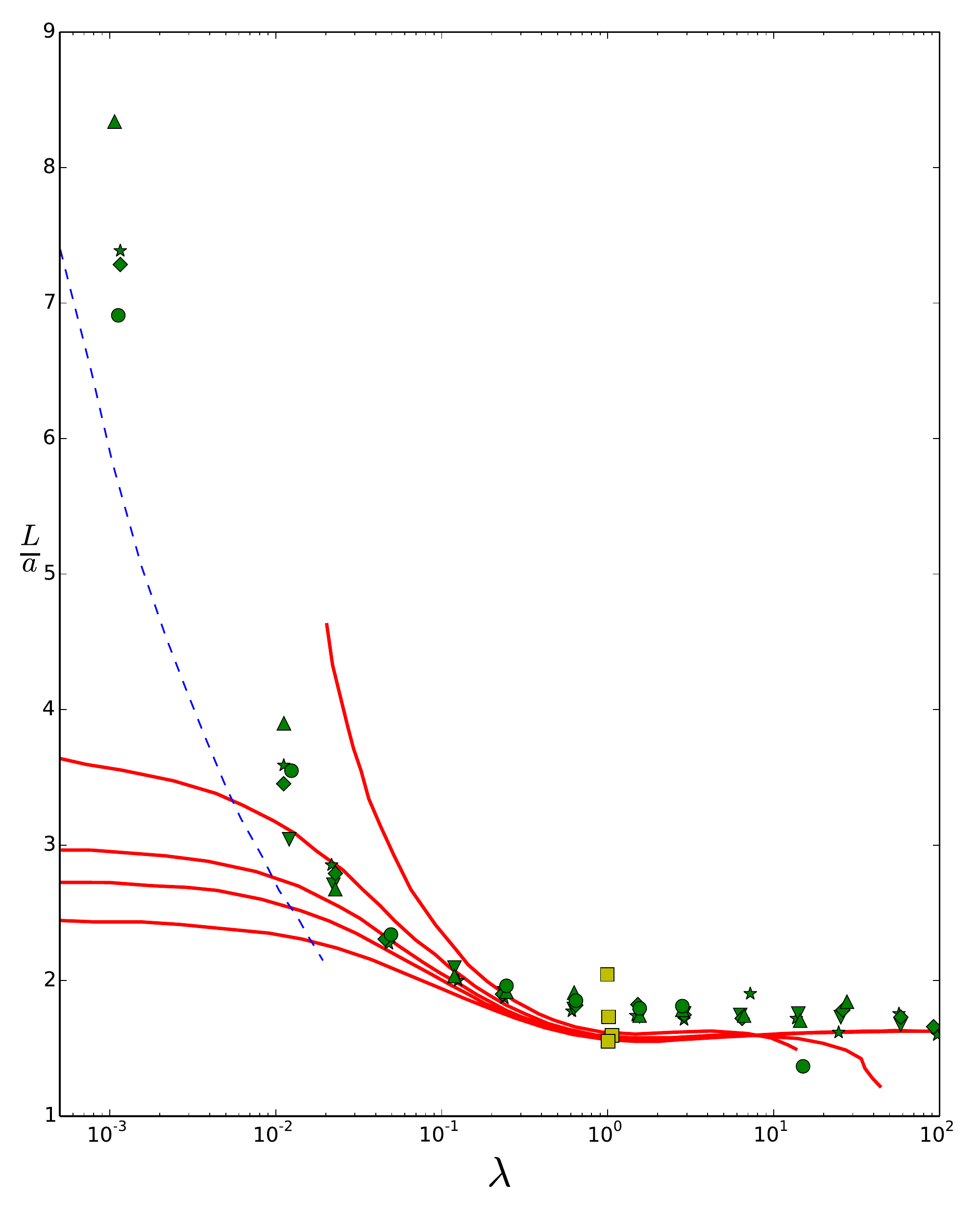}
\centering
\caption{Elongation ratio \(\frac{L}{a}\) vs viscosity ratio for different
values of \(\alpha\). Values of
\(\alpha\ (0.2,0.4,0.6,0.8\ and\ 1.0)\) increases from top to
bottom. Solid line: small-deformation theory, Dashed line:
large-deformation theory (same for all values of \(\alpha\)),
Green circles: \(\alpha = 0.2\), Green upward triangles:
\(\alpha = 0.4\), Green diamonds: \(\alpha = 0.6\),
Green downward triangles: \(\alpha = 0.8\), Green stars:
\(\alpha = 1.0\), Yellow squares: results obtained by Rallison
(1981) (values for \(\alpha = 0.8\) and
\(\alpha = 1.0\) coincide) {[}Redrawn from the Figure 29 in
Bentley and Leal (1986){]}.}
\label{fig:elongation_ratio}
\end{figure}

Bentley and Leal (1986) conducted an experimental investigation of particle deformation and breakup (Figure \ref{fig:extended_shapes}) in a simple flow field conforming to the Eq. \ref{equ:flow} with the values of \(\alpha\) = 1.0, 0.8, 0.6, 0.4, 0.2. Their most important result is the critical capillary number $\text{Ca}_{crit} = \frac{G\mu a}{\sigma}$ as a function of
viscosity ratio \(\lambda\) and the flow parameter \(\alpha\) (Figure \ref{fig:critical_capillary}). This critical value depends on the flow history (1972 Torza, 1971 Grace). Hence, the flow rate was increased very slowly so that the particle undergoes a sequence of quasi-equilibrium states and any other conditions would breakup the particle at lower rate. Breakup was said to occur at that flow rate where no stationary particle shape existed anymore. Figure \ref{fig:elongation_ratio} shows the deformation of the particles for conditions up to breakup. Finally, a reasonable agreement was achieved with small
deformation theory \((\lambda > 0.05)\) and slender body
theory\((\lambda \leq 0.01)\).

Figure \ref{fig:critical_capillary} shows the $Ca_{crit}$ number for all five flow types investigated. For high viscosity ratios (\(\lambda\) \textgreater{} 1), there is a strong qualitative dependence on the flow type which is quite
well predicted by small deformation theory (solid lines) and can be represented by an approximate relation $Ca_{crit}=0.127 \alpha^{-\frac{3}{4}} \lambda^{0.13\alpha}$. For low viscosity
ratios (\(\lambda\) \textless{} 1), the behavior is qualitatively similar for all the flow types but quantitative differences within a factor 2 to 3 are seen and can be expressed by an analytical relation $Ca_{crit}=0.1457\alpha^{-\frac{1}{2}}\lambda^{-\frac{1}{6}}$. The
ad hoc generalization of large-deformation theory (dashed lines)
predicts breakup with acceptable accuracy for the medium $\lambda$ values in the absence of theoretical data.

Figure \ref{fig:critical_capillary} and Figure \ref{fig:elongation_ratio} can be summarized as follows:

\begin{itemize}
\item
  For \(\lambda < 0.02\), the ends of the particle become
  pointed. Higher the \(\text{\ Ca}\), more is the particle elongation. Very large deformations and capillary numbers are required to cause the breakup at this $\lambda$ values.
\item
  For \(\lambda > 0.02\), the ends remained rounded up to breakup.
  Higher the viscosity ratio, smaller are the maximum stable deformation and $Ca_{crit}$.
\item
  For \(\lambda > 3\), the central portion of the particle become
  cylindrical. There exists a value of $\lambda$ beyond which the
  breakup becomes impossible and the particle deformation does not
  increase any more with increasing \(Ca\).
\end{itemize}

The steady, viscous driven breakup of the particles in simple flows is one of the earliest areas to be extensively studied and comparatively well understood. Hence, it has been accounted in brief and a detailed review would be just repetitive.

%% file: unsteady_viscous.tex
\subsection{Unsteady flow}\label{unsteady-flow}

Most of the present day research in the viscous driven breakup is directed towards understanding the phenomenon in unsteady external flows. 

\begin{figure}[t!]
\centering
\includegraphics[width=5in]{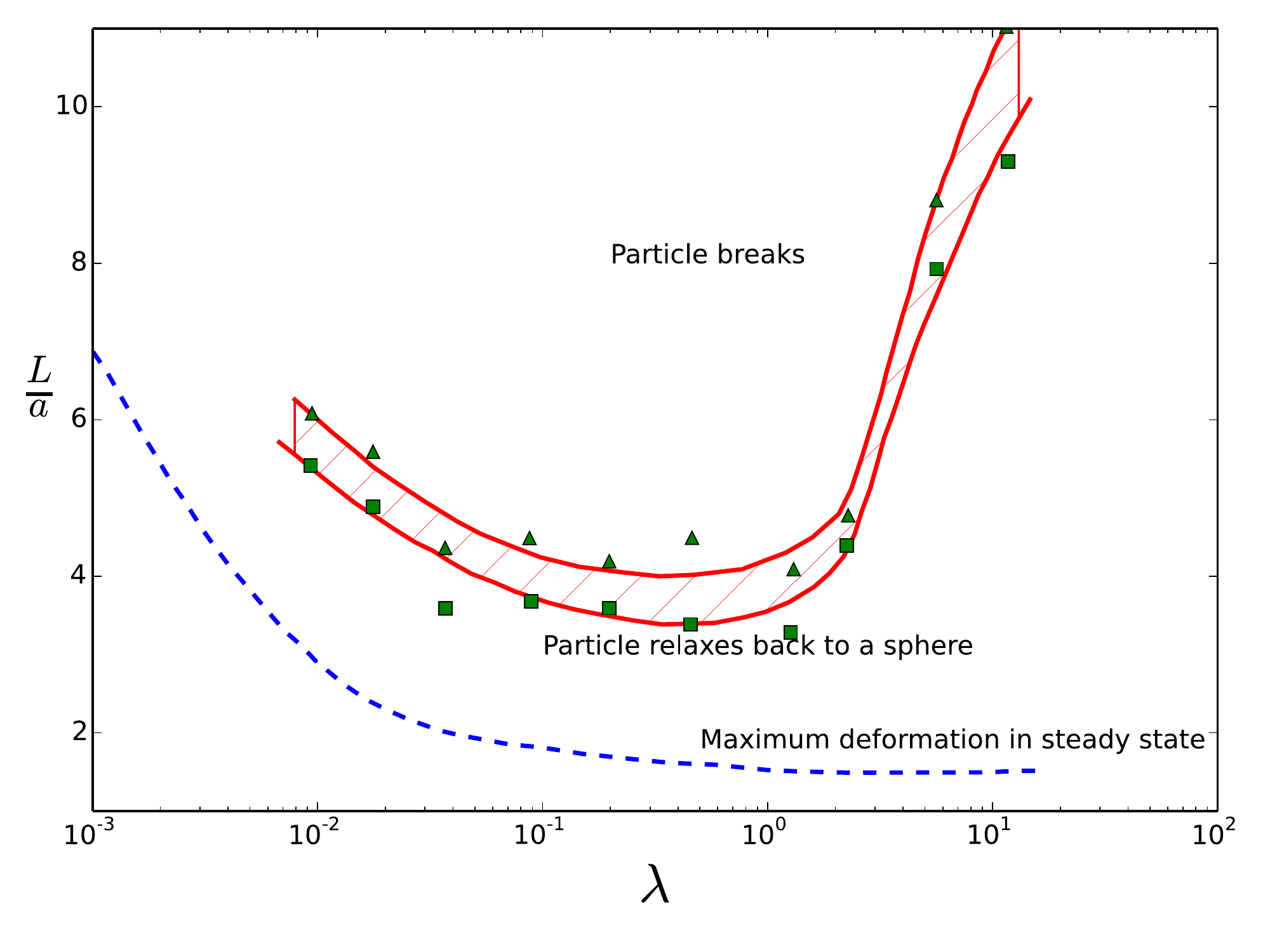}
\centering
\caption{Squares: max stable critical elongation ratio
\(\frac{L}{a}\) for the particle to relax back,
Triangles: minimum \(\frac{L}{a}\) for the particle to
break, Hatched area: shows the uncertainty in critical elongation ratio,
Dashed line: elongation ratio for the most deformed particle in steady
state observed experimentally by Bentley and Leal (1986).{[}Redrawn from
the Figure 12 in Stone, Bentley \& Leal (1986){]}.}
\label{fig:critical_elongation_ratio}
\end{figure}

In an assumed stationary case, for sufficiently small values of the $Ca$, a steady particle-shape exists in a steady two-dimensional flow
for all the values of $\lambda$. But in the majority of cases there exists a \(\ Ca_{crit}\), above which a steady particle
shape no longer exists and the viscous forces continually elongate the particle. After the particle has reached a given elongation,
if the flow magnitude is suddenly decreased to a sub-critical
value\(\ Ca_{a} < Ca_{crit}\), the particle relaxes towards the steady shape
corresponding to the final capillary number\(\ Ca_{a}\). If the initial
elongation was longer than a critical value, breakup occurs in such a
time-dependent situation. Figure \ref{fig:critical_elongation_ratio} denotes the summary of the effect of elongation ratio on the break-up of the particle.

Figure \ref{fig:critical_elongation_ratio} also shows a difficulty in particle breakup for high and low viscosity ratios. These are the observations of the experiments made by Stone, Bentley \& Leal (1986) for the case of drops. This was further extended to the case of bubbles by Kang and Leal (1987). They did
axisymmetric computations of bubbles in pure straining flow and observed
the critical elongation ratio as 2.01 at $Re_c=10$, 1.542 at $Re_c=100$ and 1.272 at $Re_c=\infty$.

\subsubsection{Capillary wave
instability}\label{capillary-wave-instability}

Plateau (1873) identified an instability in cylindrical threads due to surface tension and he proposed that a liquid cylinder under the influence of surface tension was unstable if its length was as long as its circumference. Later, Rayleigh (1879) found that a liquid jet is unstable for axial disturbances (axisymmetric disturbances) with wave numbers less than a cut-off value \(\ k_{c}\). Since the forces that are acting on the disturbed interface are the surface tension or the capillary forces, he referred
to waves as the capillary waves.

\begin{figure}[t!]
\centering
\includegraphics[width=5in]{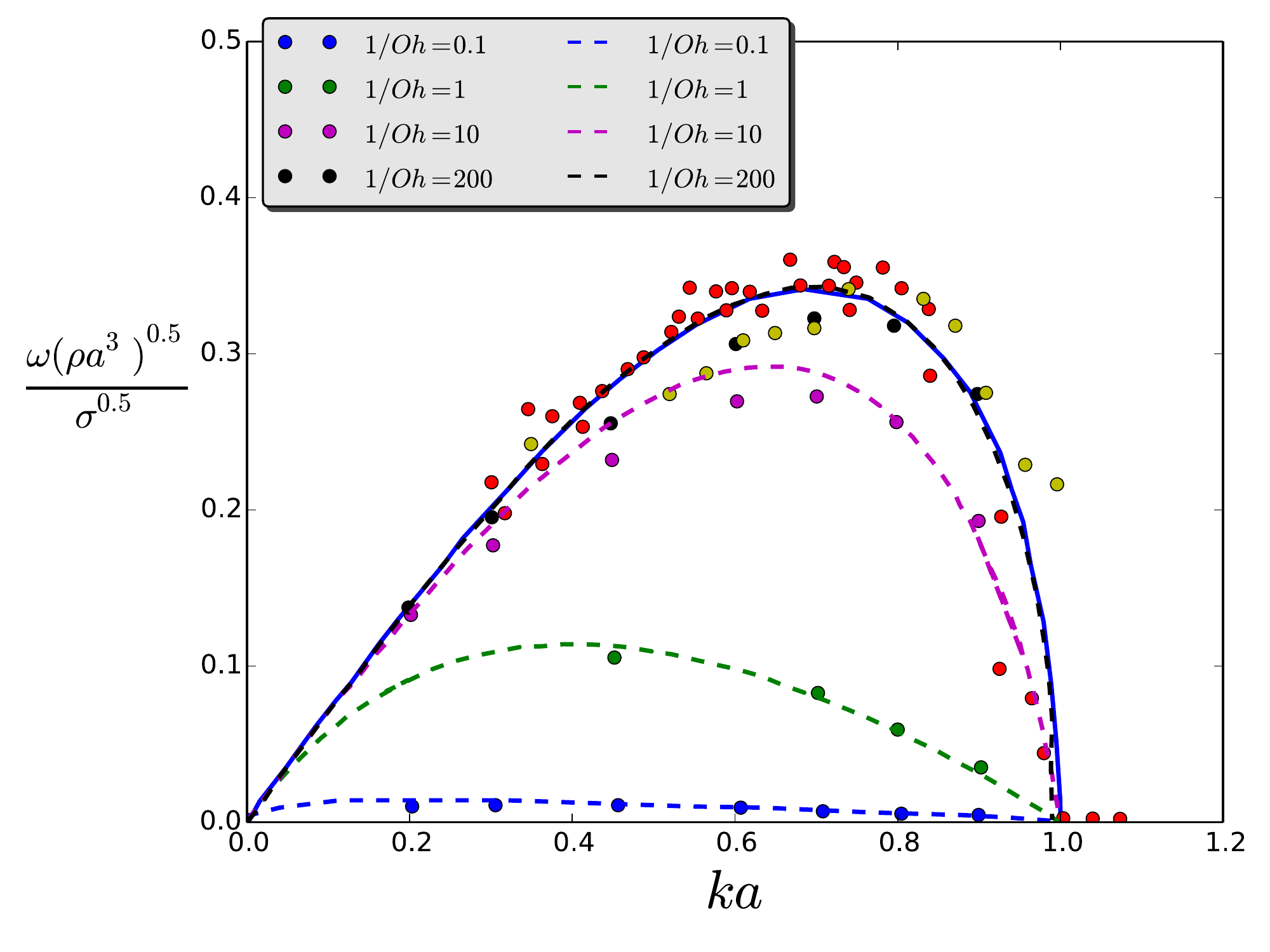}
\centering
\caption{Non-dimensional growth rate vs the wave number. Red dots:
Goedde and Yuen (1970), Blue solid-line: Theoretical result of Rayleigh
(1879), Yellow dots: Experimental data from Cline and Anthony (1978),
Blue, Green, Magenta and Black dots: Numerical results of Ashgriz and
Mashayek (1995). Blue, Green, Magenta and Black dashed-lines:
Corresponding theoretical results of Chandrashekar (1961).}
\label{fig:growth_rate}
\end{figure}

Initial theory by Rayleigh (1879) assumed an inviscid infinite liquid jet neglecting the surrounding gas effects (inviscid surrounding media) and derived a simple characteristic relationship between the wavenumber and growth rate (Figure \ref{fig:growth_rate}). The maximum growth rate occurs at $ka=0.697$. For each wavelength of unstable disturbance, one main particle and one or more usually smaller particles are formed. This theory is valid only for small disturbances acting on the interface and also provides an estimate of the daughter particle size emerging from the breakup. 

Weber (1931) and Chandrasekhar (1961) extended the Rayleigh's theory to viscous liquid jets and explained that the viscosity of the jet dampens the instability and shifts the fastest growing perturbations toward longer waves as shown in the Figure \ref{fig:growth_rate}.

Tomotika (1935) further extended the work of Rayleigh (1879, 1892) to viscous liquid threads surrounded by another viscous fluid based on the experiments in Taylor (1934). They considered both the inside and outside fluids to be at rest and derived a general equation for the relation between the growth rate $n$ and $ka$ given by,
\\
\begin{equation}
    \left| \begin{array}{cccc}
    I_1(ka) & I_1(k_{d1}a) & K_1(ka) & K_1(k_1 a) \\
    kaI_0(ka) & k_{d1}aI_0(k_{d1}a) & -kaK_0 (ka) & -k_{c1}aK_0(k_{c1}a) \\
    \frac{2\mu_d}{\mu_c}k^2I_1(ka) & \frac{\mu_d}{\mu_c}(k^2 + k_{d1}^2)I_1(k_{d1}a) & 2k^2K_1(ka) & (k^2 + k_{c1}^2)K_1(k_{c1}a)\\
    F_1 & F_2 & F_3 & F_4 \end{array} \right| = 0  
\end{equation}
\\
where $K_n(x)$ and $I_n(x)$ are $n^{th}$ order modified Bessel functions and, 
\\
\begin{equation}
\begin{aligned}
    k_{c1}^2=k^2+\frac{in}{\nu_c}\\
    k_{d1}^2=k^2+\frac{in}{\nu_d}
\end{aligned} \vast\}
\end{equation}
\\
\\
\begin{equation}
\begin{aligned}
    F_1=2i\frac{\mu_d}{\mu_c}k^2I'_1(ka)-\frac{n\rho_d}{\mu_c}I_0(ka)+\frac{T(k^2a^2-1)}{a^2n\mu_c}kI_1(ka),\\
    F_2=\frac{2i\mu_d}{\mu_c}kk_{d1}I'_1(k_{d1}a)+\frac{T(k^2a^2-1)}{a^2n\mu_c}kI_1(k_{d1}a)\\
    F_3=2ik^2K'_1(ka)+\frac{n\rho_c}{\mu_c}K_0(ka)\\
    F_4=2ikk_{d1}K'_1(k_{c1}a)
\end{aligned}
\Vast\}
\end{equation}
\\
If the inertia effects are assumed insensible  when compared to viscous effects, the equation can be reduced by applying $\rho_d\rightarrow0$ and $\rho_c\rightarrow0$.

\paragraph{Limiting solutions}\label{limiting-solutions}

When viscosity ratio was assumed to be infinity, the equation set
reduced to,
\\
\begin{equation}
    in=\frac{T(k^2a^2-1)}{2a\mu_d\Big(k^2a^2+1-\frac{k^2a^2K_0^2(ka)}{I_1^2(ka)}\Big)}
\end{equation}
\\
which is identical to Rayleigh (1892) solution. Max value of $in$ is achieved
when $ka\rightarrow0$. Therefore maximum instability when the wavelength of
varicosity is very large compared to the radius of the column.

When viscosity ratio was assumed to be tending to zero, the equation set
reduced to,
\\
\begin{equation}
    in=\frac{T(1-k^2a^2)}{2a\mu_c\Big(k^2a^2+1-\frac{k^2a^2K_0^2(ka)}{K_1^2(ka)}\Big)}
\end{equation}
\\
Similarly, even here, max value of $in$ is when $ka\rightarrow0$. Therefore, maximum
instability when the wavelength of varicosity is very large compared to
radius of the column.

For a general case, max value of $in$ was attained for $ka=0.568$, which corresponds to
a wavelength of $5.53\times2a$. They validated their theory by comparing this to a value of \(ka = 0.5\) which was obtained from experiments in Taylor (1934). For a more general
treatment of limiting cases and its applicability, refer Meister and
Scheele (1967). They also gave a generalized correlation for wave number $ka$
and the disturbance of growth rate for liquid-liquid system for a wide
range of viscosity ratio and density ratio values in Figures 6-9 in
Meister and Scheele (1967).

Another observation made by experimental studies of Taylor (1934) was
that if the apparatus were kept going, very much smaller particles were
formed than if it were stopped as soon as the initial particle had been
pulled out into a cylindrical thread, which implied a stabilizing effect
of the flow in the surrounding fluid. This concept was theoretically
analyzed by Tomotika (1936) along the similar lines of Tomotika (1935),
except that both the thread and surrounding fluid was stretched at a
uniform rate in an axisymmetric pure straining flow as in the
experiments of Taylor (1934) and inertial effects were neglected when
compared to viscous effects. They showed that the flow in surrounding
fluid limits the growth of any initial disturbance to a finite value and
proved this theoretically using the relative growth of disturbance,
wherein the value of this relative growth of amplitude of
varicosity(disturbance) was \(1:20.89\) for a cylindrical thread which
was just formed from an initial particle and \(1:2.26\times10^{10}\) for a
particle drawn into \(\frac{1}{8^{th}}\) of initial diameter.
This theory was experimentally validated by Rumscheidt and Mason (1962).
They found a good agreement for wavelengths and rates of disturbances
for the systems with viscosity ratios of 0.03 to 6.7 with the theory.
This theory was further extended by Mikami, Cox and Mason (1975) with
the help of experiments on the breakup of liquid threads in pure
straining flow.

Stone, Bentley and Leal (1986) observed that when the flow was on, there
was no evidence of capillary waves in the central cylindrical section of
less extended particles. This is because of the fact that the timescale
of end pinching mechanism (Section \ref{end-pinching}) is much smaller than the growth of capillary waves and hence the time taken by the initial infinitesimal disturbance
to reach the half cylinder is much larger than the end-pinching
mechanism. But if the particle is very highly stretched, fragmentation
occurs at the end owing to end pinching mechanism whereas capillary
waves need enough time to evolve so as to play a role in the breakup in
the central cylindrical portion of the particle.

In order to completely understand the development of capillary waves on
extended particle, Stone (1989) numerically investigated the problem of
evolution of initial disturbance. They Fourier decomposed the surface
with the isolation of the end pinching mechanism to understand the
effect of each mode. They found that the capillary wave development near
the central region of the particle was uniform in spite of the fact that
the particle was of finite length and continually shortening due to the
end pinching effect. This suggested that there was very little flow in
the central region, and hence very little effect on the capillary wave
dynamics due to continuous shortening of the particle near the ends
except for very high viscosity ratio flows where the end-effects play an
important role.

Figure 16 in Stone (1989) shows that the linear theory holds even though
the disturbance is clearly finite amplitude. Very close to actual
fragmentation, interface evolves more rapidly than theory predicts,
which demonstrates that the nonlinear effects do eventually become
noticeable, but only in the later stages of the breakup process in the
cylindrical thread-like region (formation of satellite particles).

\paragraph{Non linear theory}\label{non-linear-theory}

\begin{figure}[H]
\centering
\includegraphics[width=4in]{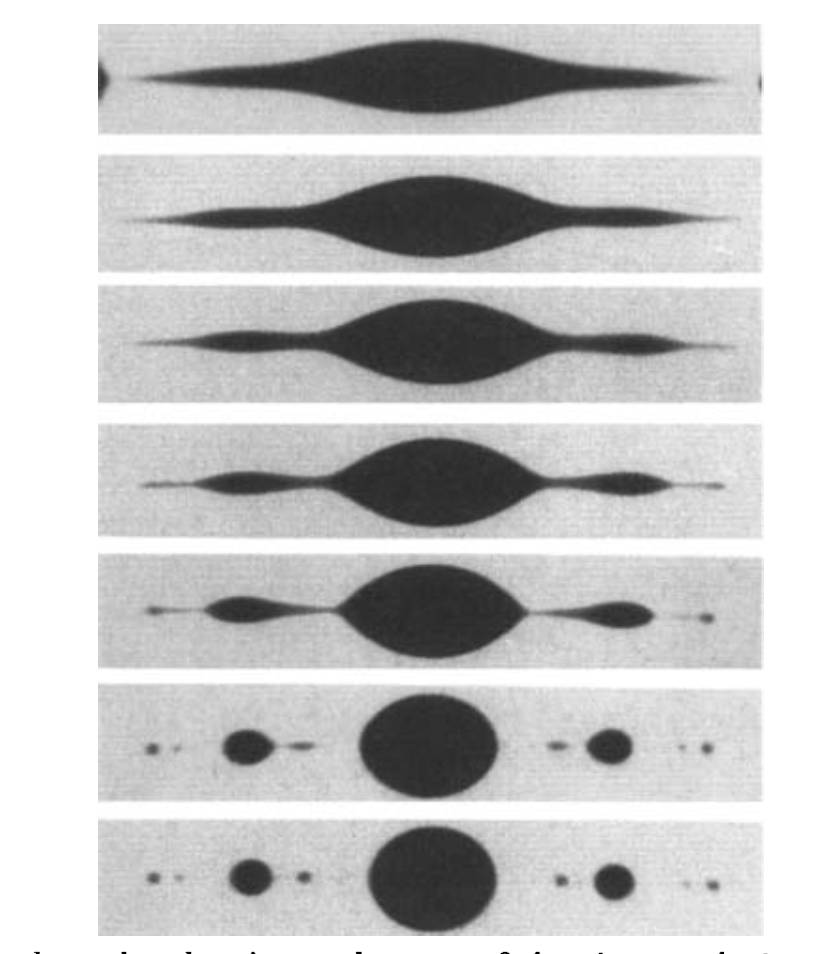}
\centering
\caption{Experimental observation of the time evolution of ligament
pinch-off from top (during first pinch-off) to bottom (during last
pinch-off) at $\lambda=0.067$ and $k=0.45$. {[}Picture taken from Tjahjadi, Stone and Ottino (1992){]}.}
\label{fig:nonlinear_theory}
\end{figure}

Linear theory is correct up to the jet breakup but cannot predict all
the sizes of the particles produced i.e it cannot predict the formation
of satellite particles, which are in turn formed from the ligaments
between the two main particles. A ligament can break into one or more
satellite particles as observed in the experiments by Tjahjadi, Stone
and Ottino (1992) (Figure \ref{fig:nonlinear_theory}). Pimbley and Lee
(1977), Ashgriz and Mashayek (1995) and Rutland and Jameson (1970) have
also done significant studies on the nonlinear theory and the formation
of satellite particles and has been reviewed extensively in Ashgriz
(2011).

\begin{figure}[H]
\centering
\includegraphics[width=5in]{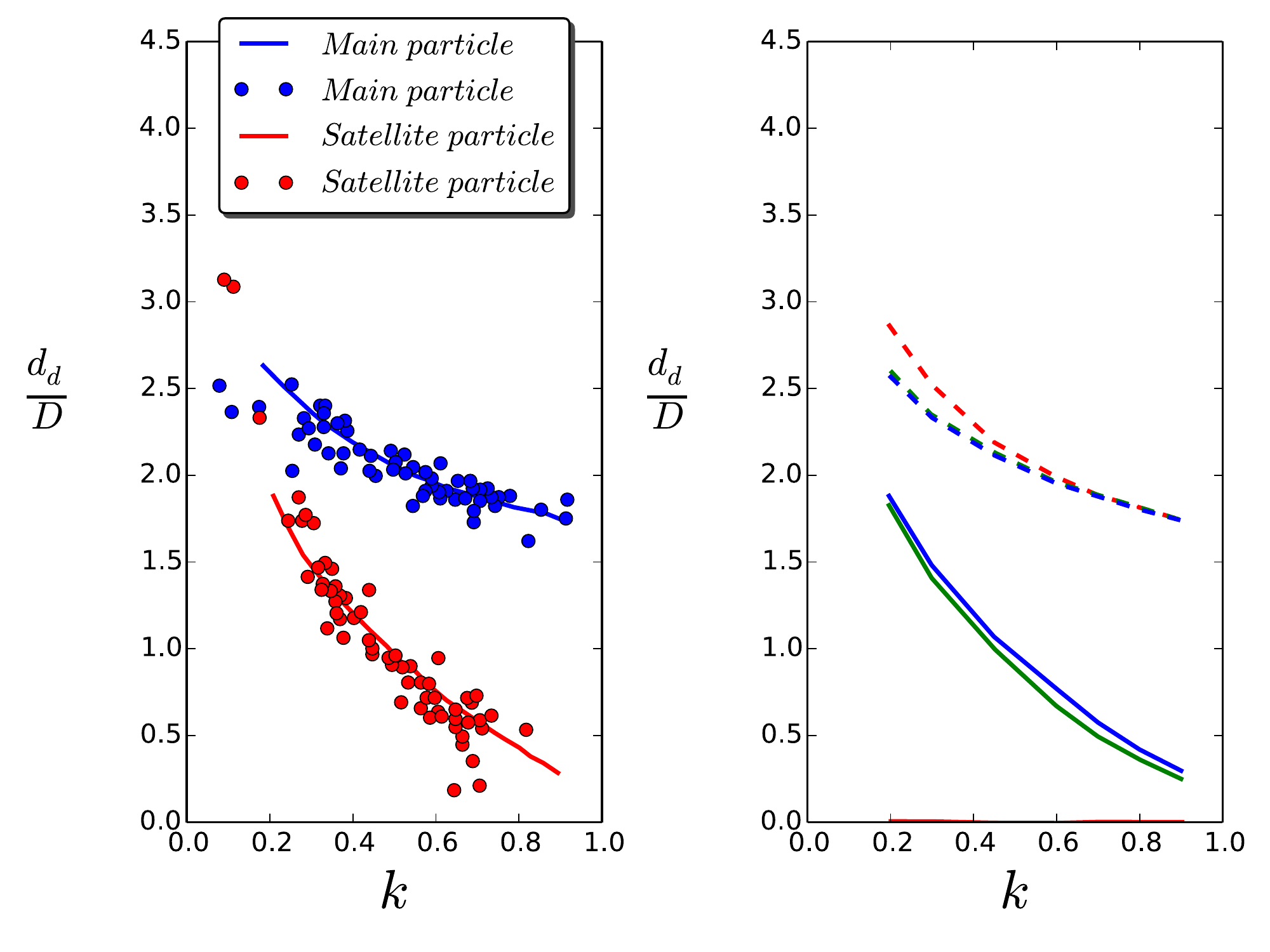}
\centering
\caption{Left Figure: Variation of size of daughter particles
($d_d$=daughter particle diameter, $D$=jet diameter), Blue and Red dots:
Experimental data of Rutland and Jameson (1970). Blue and Red lines:
Non-linear theory of Ashgriz and Mashayek (1995). Right Figure: Effect
of Ohnesorge number, Red dashed-line: Main particle size at $\frac{1}{Oh}=0.1$, Green dashed-line: Main particle size at $\frac{1}{Oh}=10$, Blue
dashed-line: Main particle size at $\frac{1}{Oh}=200$, Red solid-line: Satellite particle
size at $\frac{1}{Oh}=0.1$, Green solid-line: Satellite particle size at $\frac{1}{Oh}=10$, Blue
solid-line: Satellite particle size at $\frac{1}{Oh}=200$.{[}Redrawn from Figure 8 in
Ashgriz and Mashayek (1995){]}.}
\label{fig:daughter_size}
\end{figure}

Ashgriz and Mashayek (1995) calculated the size of the main daughter
particle and the satellite particle for different wavenumbers as shown
in the Figure \ref{fig:daughter_size}. There was no significant change in both the main
daughter particle and satellite particle size for $Oh<1$.

\begin{figure}[H]
\centering
\includegraphics[width=5in]{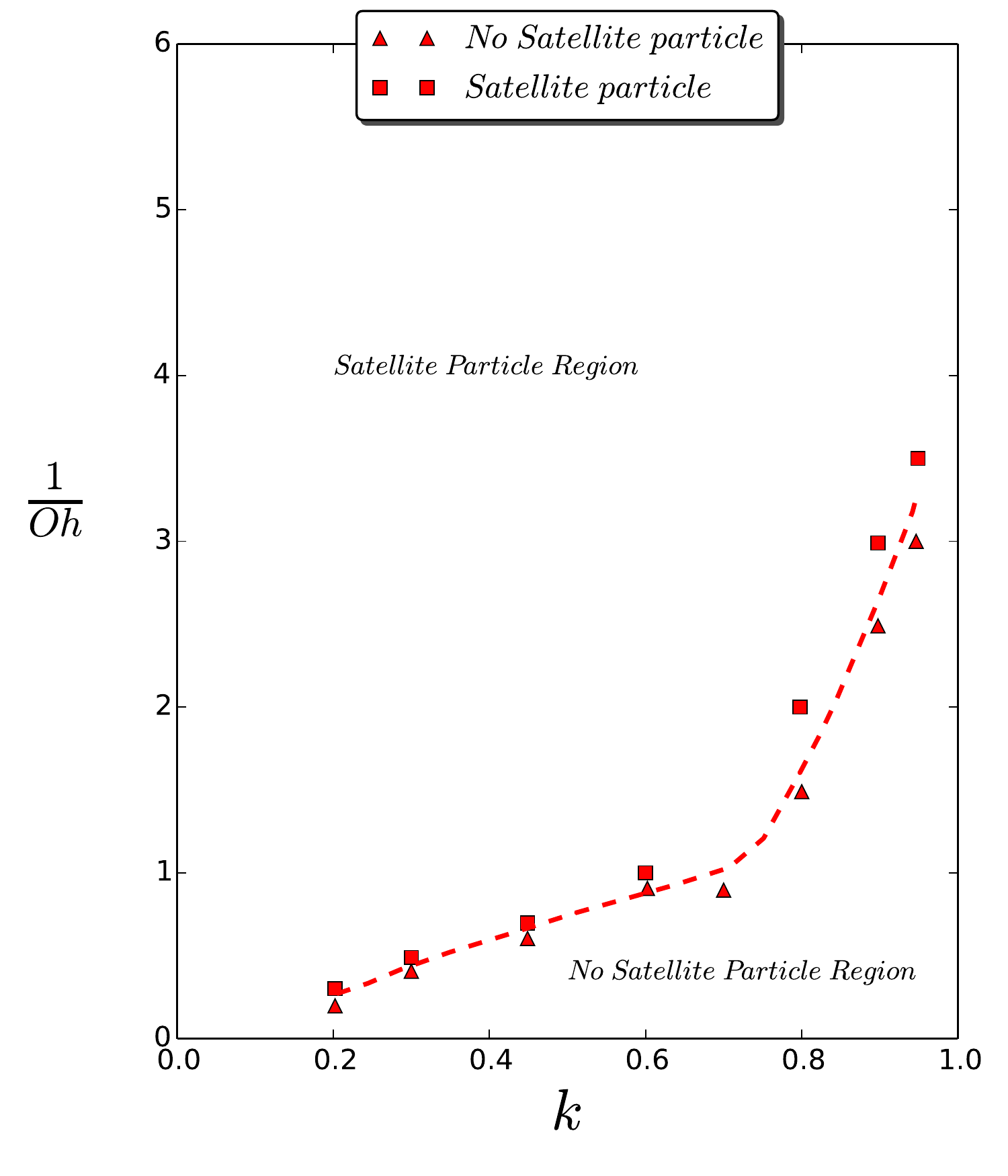}
\centering
\caption{$Oh-k$ phase diagram denoting the Satellite particle and No Satellite
particle region. {[}Redrawn from Figure 9 in Ashgriz and Mashayek
(1995){]}.}
\label{fig:oh_k}
\end{figure}

At higher values of Ohnesorge number ($Oh>0.1$), size of the satellite particles
is found to depend on the $Oh$ value. Length and diameter of the
ligament formed is found to decrease and consequently, the diameter of
the satellite particles decreases at higher values of $Oh$ and
for very high values of $Oh$, satellite particles are not
formed. They also found this limiting value of $Oh$ for
different values of $k$ as shown in the Figure \ref{fig:oh_k}. Bousfield et
al. (1986) also observed that the diameter of the satellite particles
were larger than the main particle at very low values of $k$. Further,
Spangler, Hibling and Heister (1995) studied the presence of aerodynamic
effects on the nonlinear aspects of the jet and the sizes of the main
daughter particles and the satellite particles.

Summarizing, linear capillary wave instability theory is not able to explain
the existence or formation of satellite-particles, nor is it valid for
the case when the disturbances reach finite amplitude (as observed
during breakup). It implicitly assumes that all the disturbances are
equally likely and the observed particle sizes after the breakup will
correspond to the wavelength of the fastest growing linear mode. If the
initial amplitude of this critical disturbance is known, then the time
for breakup can be estimated. In addition, wavelength and growth rate of
the critical disturbance depend on the viscosity ratio. For bubbles the
inviscid limit is a good approximation, but for large viscosity ratios
growth slows down. Whereas, nonlinear theory can explain the breakup of ligaments and the late stages of the breakup of jets, which in turn dictates the size of the satellite particles.

\subsubsection{End pinching}\label{end-pinching}

Among the cases in which a particle was initially extended and then left
free to observe its deformation, Stone (1986, 1989) found that there was no capillary
waves during the elongation process or when the flow was on. There was
also no such case wherein the extending particle fractured at the
central region when the flow was on but the breakup was seen when the
particle was left free. They observed that this breakup
was due to the capillary pressure variations near the ends rather than
the instability of the infinitesimal disturbances in particle shape and hence, named it as "end-pinching" mechanism.

For low viscosity particles, large interior velocity gradients are
possible in the cylindrical region due to the pressure gradient unlike
the high viscosity particles. This is the main cause for the end
pinching mechanism.

When the low viscosity particles are subjected to pure straining flow,
they develop long slender shapes with pointed ends. After the flow is
turned off, the high initial curvature at the ends results in the large
velocities near the ends and a very rapid reduction in the initial
length. The initial pressure gradient drives the rapid relaxation and as
the end becomes more spherical this pressure gradient diminishes. The
pressure (due to the local curvature of the interface) also generates a
local flow from the cylindrical region which thus causes a neck to form
in the particle shape as illustrated in Figure 9 and also in the Figure 6 of
Stone and Leal (1989).

Therefore, the mechanism for relaxation and breakup of an extended
particle in an otherwise quiescent fluid consists of a competition
between a pressure-driven flow near the end, which causes translation of
the end toward the particle centre (thus tending to return the particle
to its spherical equilibrium shape), and a pressure-driven flow away
from the centre in the transition region which leads to the development
of a neck and thus to breakup via a capillary pinch-off process.

\begin{figure}[t!]
\centering
\includegraphics[width=5in]{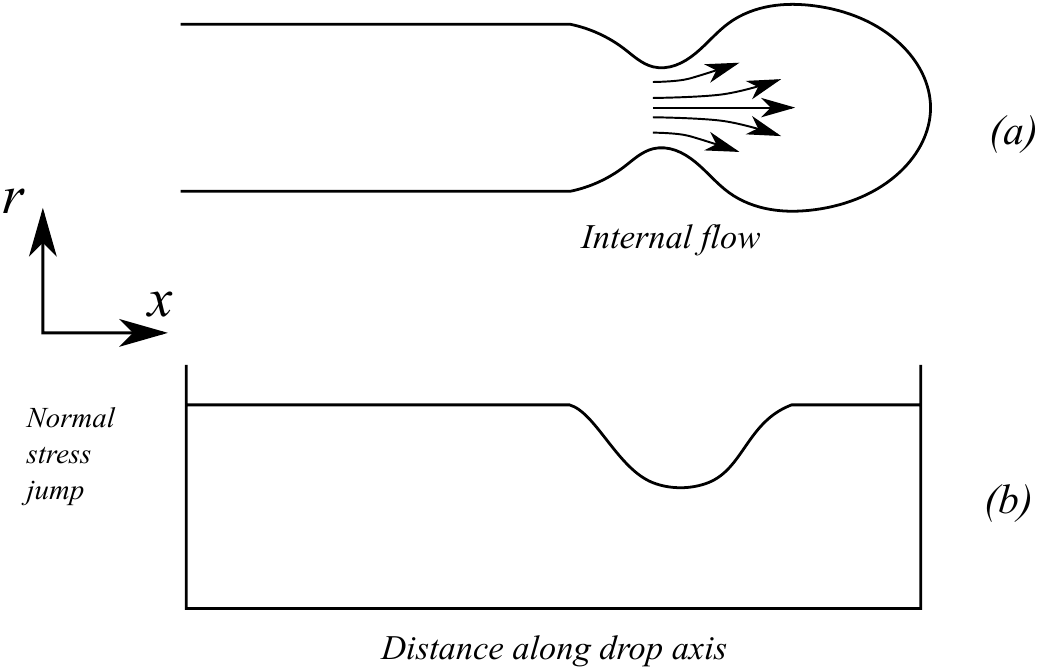}
\centering
\caption{(a) Typical particle shape shortly after the flow is stopped.
(b) Approximate normal stress jump across the interface.}
\label{fig:end_pinching}
\end{figure}

Stone, Bentley and Leal (1986) also suggested that the final remnant
parent particle size is determined by the rate at which ends bulb up and
contract towards the particle centre, relative to the rate at which the
ends pinch off. The timescale for this mechanism is much smaller than
the growth of capillary waves.

Recently there have been few numerical studies (Li, Renardy and Renardy
(2000), Renardy and Cristini (2001a,b) and Renardy, Cristini and Li
(2002) using VOF, Komrakova et al. (2015) using diffuse interface free
energy LBM) that observe the end pinching mechanism. Komrakova et al.
(2015) observed end pinching mechanism at low shear rates and $\lambda$ low values
and observed capillary wave instability at low shear rates and high $\lambda$
values.

\subsubsection{Tip streaming and Tip
dropping}\label{tip-streaming-and-tip-dropping}

\begin{figure}[H]
\centering
\includegraphics[width=3.5in]{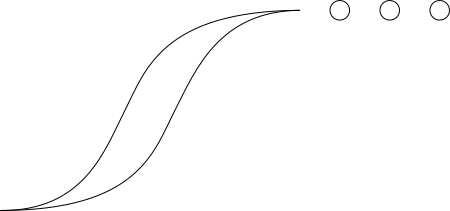}
\centering
\caption{Graphical representation of Tip Streaming phenomenon.}
\label{fig:tip_streaming}
\end{figure}

Tip streaming and tip dropping are the modes of particle breakup that occur
when surfactants are present. Taylor (1934) was the first to report the
tip streaming phenomenon and he referred to it as a transient phenomenon
since it disappeared when the shear rate was further increased. The true
cause was identified by De Bruijn (1989, 1993). Janssen, Boon and
Agterof (1994, 1997) were the first to observe the tip dropping
phenomenon and also observed tip streaming in both simple shear flow and
pure straining flows. For these breakup mechanisms in simple shear flows
the particle develops a sigmoidal shape and a stream of tiny particles
are ruptured off the tips of the parent particle as shown in Figure \ref{fig:tip_streaming}. This
mechanism occurs for capillary numbers less than the critical value and
it is important because the shear rates required for this type of
breakup have in some circumstances been observed to be two orders of
magnitude lower than that for the normal breakup (clean particle) in
which the particle is broken in two or three almost equally sized
particles with a few tiny satellite particles in between.

According to De Bruijn (1989, 1993) tip streaming results from
interfacial tension gradients with lower value at the tip and higher
elsewhere. This phenomenon occurs when there is a moderate level of the
surface active material. At lower levels of the surface active
materials, surface tension is not low enough at the tip and at higher
levels, surface tension is lowered uniformly and the gradients are too
weak. Such interfacial tensions gradients make the particle surface less
mobile allowing the shear stresses exerted by the continuous phase to
pull out a stream of tip particles. This explanation however can only
hold good when the shear stresses exerted by the continuous phase are
large enough to maintain such an interfacial tension gradient and when
the diffusion of the surface active material from the particle to the
surface is slow enough not to interfere with the buildup of the
interfacial tension gradients.

De Bruijn (1989, 1993) experimentally determined the effect of various
parameters on the tip streaming mechanism and found that it depends on
the fluids used (doesn't seem to occur in all fluids) irrespective of
the value of surface tension provided that the viscosity ratio is very
less than one. They also found that this mechanism is some sort of
depletion mechanism effect and depends on the history of the particle.
The ruptured particle radii were two orders of magnitude smaller than
their parent particle and had significantly reduced surface tensions.

Eggleton, Tsai and Stebe (2001) numerically simulated the particle
breakup in linear uniaxial pure straining flow using a nonlinear model
for the surface tension. They explained that the surface convection
sweeps surfactants to the particle poles, where it accumulates and
drives the surface tension to near zero. The particle assumes a
transient shape with highly pointed tips. From these tips, thin liquid
threads are pulled. Subsequently, small, surfactant-rich particles are
emitted from the termini of these threads. After a finite number of
daughter particles have been ejected, the cleaner parent particle can
attain a stable shape. The scales of the shed particles are much smaller
and depends on the initial surfactant coverage.

\begin{figure}[H]
\centering
\includegraphics[width=3.5in]{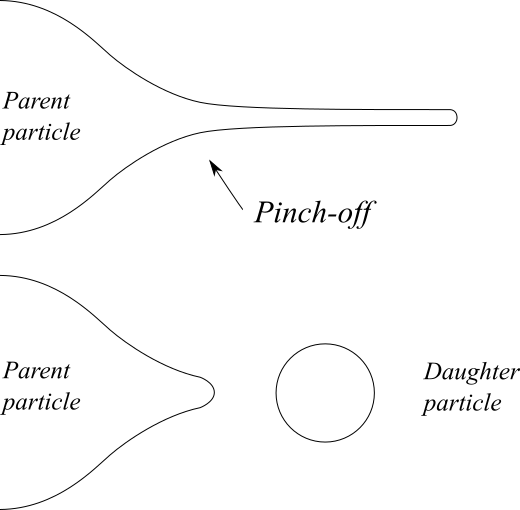}
\centering
\caption{Graphical representation of Tip Dropping phenomenon.}
\label{fig:tip_dropping}
\end{figure}

Dilute initial coverage leads to tip streaming, while high initial
coverage leads to another mode of breakup called as tip dropping,
wherein the daughter particles are larger and ejected more
intermittently. The mechanism which causes the liquid threads to be
pulled is same as that in the tip streaming phenomenon but due to the
high initial surfactant coverage, larger liquid thread is pulled and the
pinching takes place at the junction of the parent particle-thread as
shown in Figure \ref{fig:tip_dropping}, which eventually forms a larger daughter particle.

Numerical simulations in Eggleton, Tsai and Stebe (2001) also showed
that at trace surface concentrations, surface tension effects are
negligible and the critical strain rate for a clean interface is reached
before $(\Gamma_{\infty}$ is approached at the pole. The particle
becomes unstable and extends as a continuous thread, with the
possibility of surfactants altering the breakup. At dilute
concentrations convection drives the surfactants towards the pole and
$\Gamma_{\infty}$ is reached at strain rates below the critical strain
rate for the clean particle. Surface tension at the pole approaches zero
and sharp gradients in surface tension lead to the emission of a thread.
Thread size increases with concentration. If significant surface tension
gradients are present at high concentrations, the thread diameter will
be comparable to the parent diameter, leading to particle fracture. Tip
streaming occurs only when the $\lambda < 0.2$.

%% file: inertial.tex
\section{Inertial force driven
breakup}\label{inertial-force-driven-breakup}

As the Reynolds number increases, the Stokes flow approximation is no
more valid and the inertial forces begin to dominate over the viscous
forces. Hence, Weber number, Reynolds number, density ratio and
viscosity ratio are the important parameters used to characterize the
flow field. But depending on the viscosity, of the dispersed particles,
Ohnesorge number is also considered for the study.

\subsection{Straining flows}\label{straining-flows}

Misksis (1981) numerically calculated the steady particle shape in an
axisymmetric pure straining flow in the limit of infinite Reynolds
number using the boundary-integral technique. They found that there is
no steady shape beyond a Weber number of value 2.76 which was called a
critical Weber number. They were the first to introduce this concept of
critical Weber number. Later, Ryskin and Leal (1984) compared and
extended the work of Misksis (1981) by considering the deformation of a
particle in a uniaxial pure straining flow at finite Reynolds number
values (\(0.1 \leq Re \leq 100\)). Critical Weber number was defined
based on the observation of appearance of the waist and subsequent
divergence of their numerical scheme signifying particle breakup. They
also stated that the result provided by Misksis is accurate for
\(Re \geq 100\) and provided an interpolation formula for lower values
of Reynolds number.
\\
\begin{equation}
    \Big(\frac{1}{We_{crit}}\Big)^\frac{10}{9}=\Big(\frac{1}{2.76}\Big)^\frac{10}{9}+\Big(\frac{1}{0.247Re^\frac{3}{4}}\Big)^\frac{10}{9}
\end{equation}
\\
They stated that the above case is only valid for bubble and not for
inviscid drops and the two cases is identical only if the Reynolds
number is zero or if the shape is fixed (spherical). Otherwise, one must
take into account the variation of pressure inside the inviscid drop.

\begin{figure}[t!]
\centering
\includegraphics[width=5in]{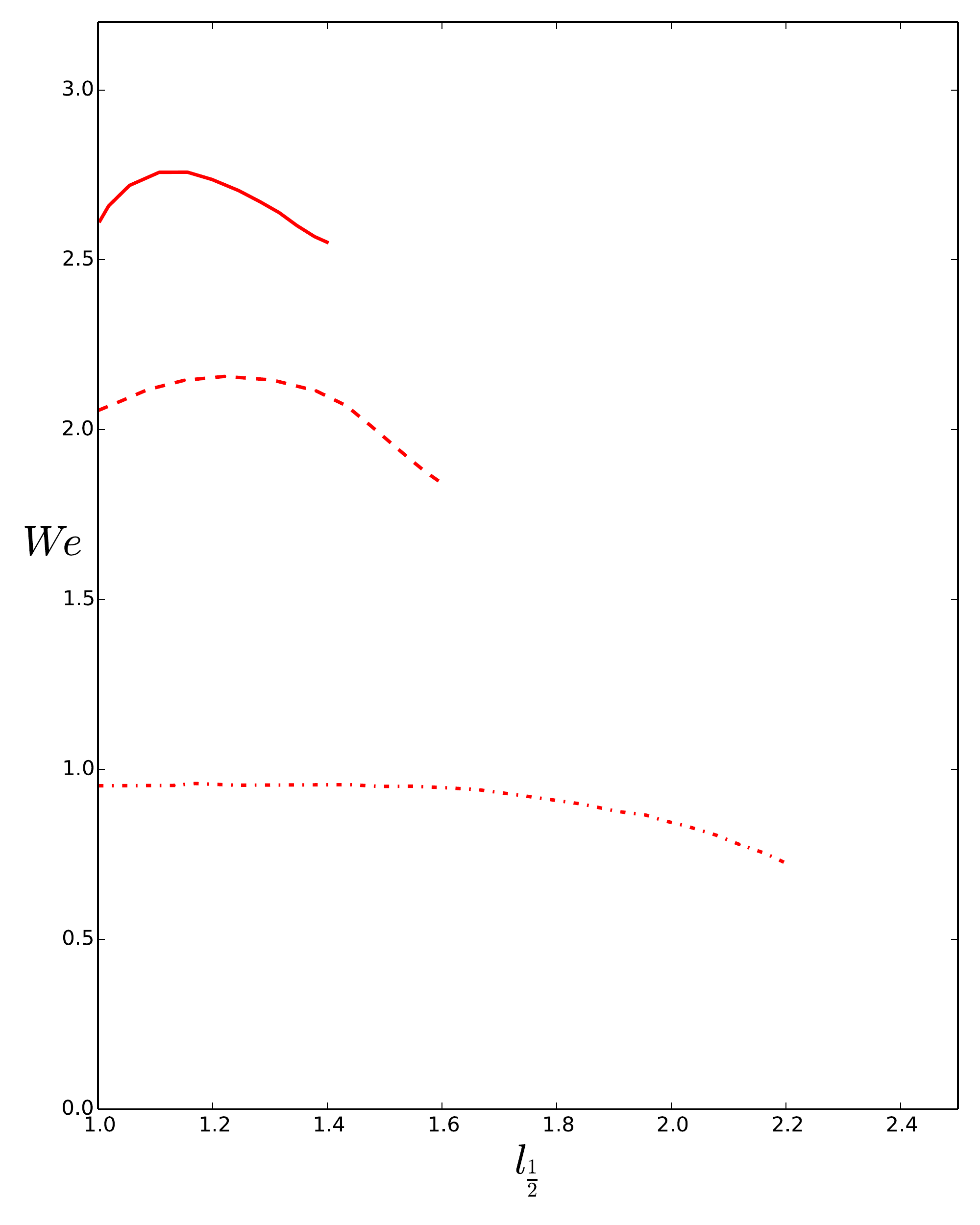}
\centering
\caption{Effect of initial half-length of the particle
\(l_{\frac{1}{2}}\) on the critical Weber
number. Solid line: \(Re = \infty\), Dashed line:
\(Re = 100\), Dotted line: \(Re = 10\) {[}Redrawn from
Figure 7 in Kang and Leal (1987){]}.}
\label{fig:straining_flow}
\end{figure}

Kang and Leal (1987, 1989) studied the steady and unsteady deformation
of a particle in a uniaxial and biaxial pure straining flow. They
attempted to correlate the initial half-length \(l_{1/2}\)with the
critical Weber number (defined based on the steady state calculations)
and observed (Figure \ref{fig:straining_flow}) that it decreases as the initial elongation
from steady shape increases and stated that the particle shape in a
subcritical flow is stable in the neighborhood of steady state and
unstable if sufficiently deformed at some initial instant as a
consequence of stretching in a flow with
\(We < \text{We}_{\text{crit}}\). A limit point observed for the
existence of steady axisymmetric particle shapes at a Weber number of
\(O(2 - 3)\) in uniaxial flow for all\(\ Re \geq 10\) was not found for
biaxial flow at any \(We \leq O(10)\) for $Re$ up to 200.

Other works on biaxial pure straining flows are Frankel and Acrivos
(1970) who calculated particle shape in creeping flow limits to moderate
values of capillary number\((Ca \leq 0.2)\) and Hinze (1955), Lewis and
Davidson (1982) at higher Reynolds number. Further uniaxial pure
straining flow breakup is assumed to be a simplified model for breakup
of particles in turbulent flows. Hence more studies on pure straining
flows has been covered in the Section \ref{instantaneous-breakup} (\textbf{Instantaneous
breakup)}.

\subsection{Secondary breakup}\label{secondary-breakup}

Accelerating particles are found ubiquitous in the nature. Importance of
these have been found in the wide range of applications ranging from
rainfall in the nature to the spray atomization in combustion engines.
Considering the vast research on this field and also accounting for the
large number of existing reviews, a very brief account has been
considered here.

When a particle breaks apart into a multitude of small fragments due to
disruptive aerodynamic forces, the process is termed as secondary
atomization. Pilch and Erdman (1987) defined the total breakup time as
the time taken for a particle and its fragments to reach the steady
state. Multiple correlations have been proposed and some are not in
agreement with others refer Pilch and Erdman (1987), Hsiang and Faeth
(1992), Gelfand (1996) and Guildenbecher, López-Rivera and Sojka (2006)
for the correlations.

The dominating forces that govern the breakup at high Reynolds number
and high density ratio are aerodynamic forces which are counteracted by
the restorative surface tension forces. The radical mode of breakup is
found to be different for different magnitude of these competing forces.
Hence the breakup mechanism is divided into different breakup regimes
based on the range of aerodynamic Weber numbers (most important
parameter for secondary atomization) as shown in the Table \ref{tab:transitional_we} at lower
values of particle Ohnesorge number and higher values of Reynolds
number. At higher values of Ohnesorge number $(Oh > 0.1)$, particle
viscosity hinders deformation and also dissipates energy supplied by the
aerodynamic forces, which reduces the tendency towards breakup, and
hence was found to change the transitional $We$ values to
some extent. Many correlations are present for $We_{crit}$
(weber number at which bag breakup starts) at high Oh but none is known
to be accurate at $Oh > 1$. Reynolds number does not affect
the breakup process, though at lower values , Aalburg, van Leer and
Faeth (2003) has numerically observed significant change in transitional
values. Han and Tryggvasson (1999, 2001) and Aalburg, van Leer and Faeth
(2003) also observed that the critical Weber number increases as the
density ratio reduces to 1 and is independent above 32.

\begin{longtable}[]{@{}ll@{}}
\toprule
\textbf{Regime} & \(We\) Range\tabularnewline
\midrule
\endhead
Vibrational (no breakup) & \(0 < We < \sim 11\)\tabularnewline
Bag breakup & \(\sim 11 < We < \sim 35\)\tabularnewline
Multimode breakup & \(\sim 35 < We < \sim 80\)\tabularnewline
Sheet thinning breakup & \(\sim 80 < We < \sim 350\)\tabularnewline
Catastrophic breakup & \(We > \sim 350\)\tabularnewline
\bottomrule
\caption{$We$ for particles with $Oh<0.1$ [Adapted from Guildenbecher, Lopez-Rivera and Soika (2006)].}
\label{tab:transitional_we}
\end{longtable}

\subsubsection{Deformation and vibrational breakup}\label{deformation-and-vibrational-breakup}

When a spherical particle enters the disruptive flow field, due to the
higher static pressure at front and rear stagnation points when compared
to the particle periphery, it deforms into an oblate spheroid. In some
cases, the particle oscillates due to restoration by surface tension and
this oscillation might lead to breakup. According to Pilch and Erdman
(1987) this breakup mode does not lead to small final fragment sizes and
the breakup time is long compared to other breakup modes.

\subsubsection{Bag breakup}\label{bag-breakup}

The \(\text{We}\) number at which the bag breakup starts is termed as
the critical Weber number, $We_{crit}=11\pm2$ (Guildenbecher 2009). In this mode
of breakup, a bag is formed due to the pressure difference between the
leading stagnation point and the wake of the particle (Han and
Tryggvason 1999, 2001), though the exact reason for the formation of bag is still an open question. The outer edge forms a toroidal ring to which
the bag is attached (Figure \ref{fig:secondary_breakup}a). Chou and Faeth (1998) also reported
that bag breakage results in larger number of small fragments (average
size 4\% of parent particle), whereas the toroidal ring into smaller
number of large fragments (average size 30\% of parent particle).

Chou and Faeth (1998) studied the temporal characteristics of this
breakup and showed that it requires $5\sim6$ characteristic timescale $t_c$
for complete breakup, where $t_c$ is given by,
\\
\begin{equation}
    t_c=\frac{D_0\Big(\frac{\rho_d}{\rho_c}\Big)^\frac{1}{2}}{U}
\end{equation}
\\

\subsubsection{Multimode
breakup/ transition or chaotic
breakup}\label{multimode-breakup-transition-or-chaotic-breakup}

\begin{figure}[H]
\centering
\includegraphics[width=3in]{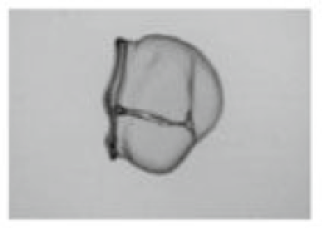}
\centering
\caption{Multimode breakup {[}Picture taken from Guildenbecher and
Sojka (2011){]}.}
\label{fig:multimode}
\end{figure}

It is a combination of two breakup modes or can be considered as a
transition between two breakup modes. As seen in the Figure \ref{fig:multimode}, it is
accompanied by a bag and a long filament at the center, which is also
referred as plume/stamen (Pilch and Erdman 1987). It can also be a
combination of bag and sheet-thinning breakup (Dai and Faeth 2001) or a
multi-bag breakup (Cao et al. 2007) (Figure \ref{fig:secondary_breakup}c).

Rayleigh-Taylor instability was thought to be the mechanism behind this
breakup along with the aerodynamic intensification effects and hence was
called a combined "Rayleigh-Taylor/aerodynamic drag" mechanism (Jain et al. 2015, Jain et al. 2016, 2017).

\subsubsection{Sheet-thinning
breakup}\label{sheet-thinning-breakup}

\begin{figure}[H]
\centering
\includegraphics[width=5in]{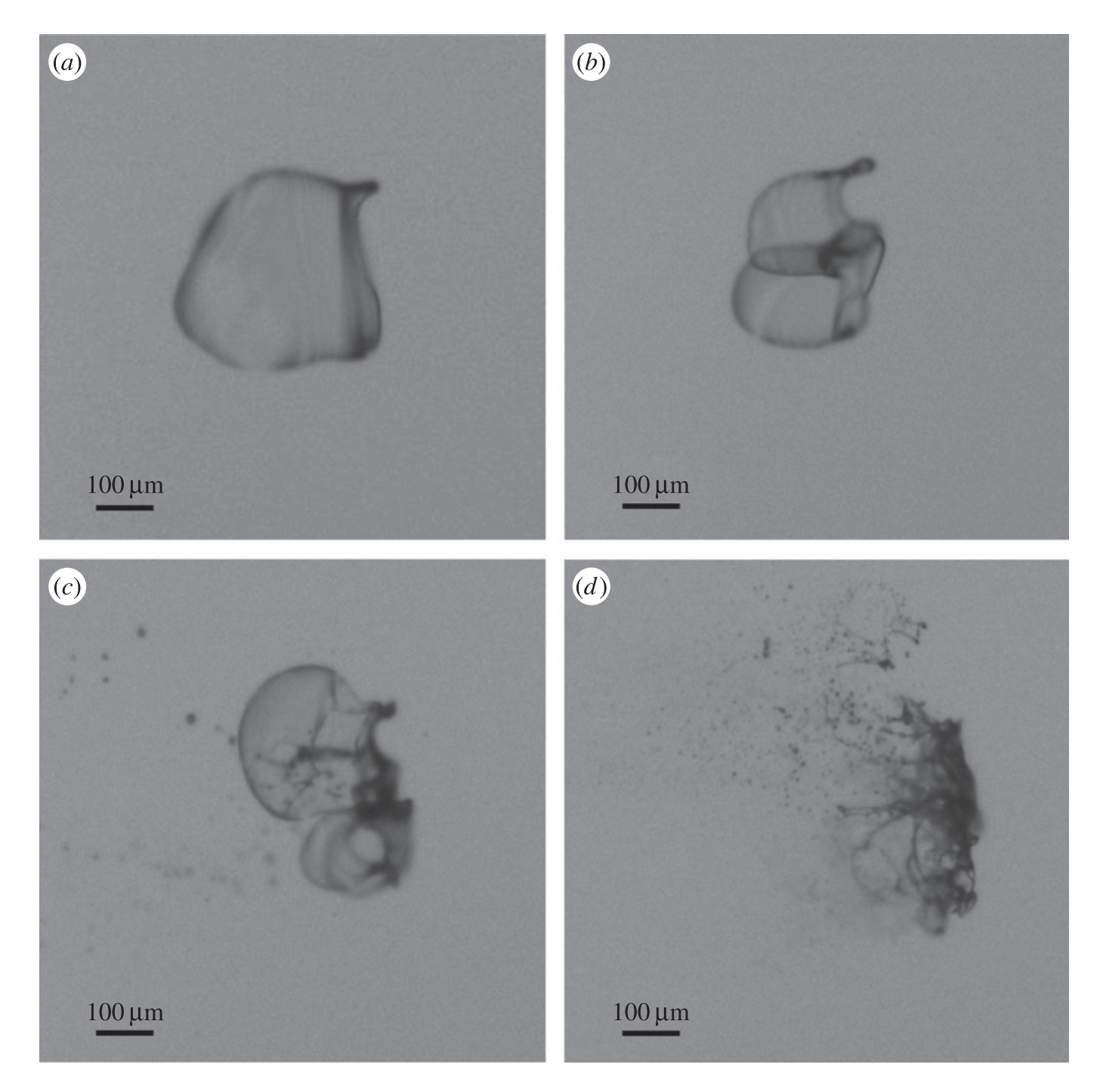}
\centering
\caption{Secondary breakup of particles. (a) \(We = 20\) (bag
breakup), (b) \(We = 40\) (bag-stamen breakup), (c)
\(We = 80\) (multi-bag breakup) and (d) \(We = 120\)
(sheet-thinning breakup). {[}Picture taken from Jain et al. (2015){]}.}
\label{fig:secondary_breakup}
\end{figure}

According to Liu and Reitz (1997), at higher \(We\) numbers a
sheet is formed at the deflected periphery of the particle which evolves
into ligaments that break up into multiple fragments. This process
continues until the particle is completely fragmented or until it has
accelerated to the point at which aerodynamic forces are negligible and
finally a core particle remains at the end (Figure \ref{fig:secondary_breakup}d). This process
was mistakenly assumed to be due to the shear from the continuous phase
flow over the deformed particle which results in the formation of the
boundary layer inside the particle surface and this boundary layer
becomes unstable at the periphery resulting in the stripping of mass and
hence was previously called as Shear Stripping (Ranger and Nicholls
1969).

\subsubsection{Catastrophic
Breakup/piercing
breakup}\label{catastrophic-breakuppiercing-breakup}

At very high \(We\) numbers, unstable surface waves have been
observed on the leading edge of the particle, which eventually
penetrates into the particle causing multiple fragment breakup. Liu and
Reitz (1993) considered this phenomenon to be a result of
Rayleigh-Taylor instability due to the acceleration of lighter medium
towards the denser medium. Though, most experimental observations show
the formation of waves on the front end of the particles, there has been
few speculations on the occurrence of the waves on the rear end of the
particles.

Recently there have been few numerical studies of secondary breakup.
Works that study secondary breakup at low density ratios $(\frac{\rho_d}{\rho_c}<100)$ are Kékesi, Amberg and Wittberg (2014), Han
and Tryggvason (1999) and at high density ratios by Xiao, Dianat and
McGuirk (2014) and Jain et al. (2015). Renardy (2008) and Kékesi (2016)
studied the breakup when both the inertia and viscous forces are
considerable i.e., inertial breakup with shear. Renardy (2008) studied
the inertial breakup of a particle in an initially established shearing
external fluid at subcritical capillary numbers. They observed that the
particle extended more than when it was placed in a gradually increasing
shearing external fluid before it broke. Kékesi 2016 in their numerical
simulations observed 5 different modes (bag, shear, jellyfish shear,
thick rim shear, thick rim bag) of breakup in the presence of inertia
and shear. Increasing the Reynolds number, shear rate and decreasing
viscosity ratio they observed the breakup mode tending towards the
shearing breakup.

At very low density ratios, transitional $We$ has not been found to match
with the experiments. Hence in a recent study by Jain et al. (2016, 2017), the effect of density ratio on the breakup was studied and it was found that there exists a critical density ratio value below which the dynamics is different from that at the higher values(observed in experiments). Though most of the numerical and experimental studies try to understand the mechanism behind the different breakup
modes, a good agreement between the different studies has not been
achieved yet and hence there is a huge scope for better and more
detailed numerical and experimental studies.

\subsection{Breakup under the action of gravity in stagnant
media}\label{breakup-under-the-action-of-gravity-in-stagnant-media}

Particles under the action of gravity involves freely rising or falling
particles in a stagnant medium. The breakup mechanism in this situation
is explained as follows: the breakup can only occur if the disturbance
located at the advancing front of the particle grows quickly enough and
reaches sufficient amplitude before being swept around the equator.
Ryskin and Leal (1984) compared the deformation of a rising particle in
quiescent fluid to the pure straining flow and stated that the
phenomenon is much simpler in case of pure straining flow. In the case
of rising particle, unlike the pure straining flow, capillary number
does not play a role even at the low Reynolds numbers. Though, the Weber
number does determine the deformation at non-zero Reynolds numbers, the
deformation is rather small at \(We = O(1)\), and, indeed extremely high
Weber numbers can be reached experimentally for spherical-cap particles
without breakup, the shape of a particle becoming essentially
independent of \(We\) above some value of order 15-20. Hence
neither \(Ca\) or \(We\) can be a direct measure of the
strength of the dominant deforming force relative to the restoring
tendency of the surface tension.

Earliest experimental work on the breakup of particles freely moving
under the action of gravity in liquid-liquid systems can be found in Hu
and Kinter (1955).Klett (1971), Ryan (1978) analyzed liquid drops falling in
air. Other works which focused on bubbles are Clift and Grace
(1973), Clift, Grace and Weber (1974). Krishna, Venkateswarlu
and Narasimhamurty (1959), correlated the maximum size of a liquid drop
falling in water.
\\
\begin{equation}
    d_{max}=\frac{2.25U_M^2}{g(\rho_d-\rho_c)}\ in\ cm
\end{equation}
\\
where \(U_{M}\) is in \(cm/s\) and is given by,
\\
\begin{equation}
    U_M=\frac{0.568\sigma}{\mu_c S_d^{1.305}}
\end{equation}
\\
and \(S_{d}\)is given by,
\\
\begin{equation}
    S_d=\frac{g_c\sigma}{\mu}\Big(\frac{3}{4g}*\frac{\rho^2}{\mu_c \Delta\rho}\Big)^{1.3}
\end{equation}
\\
where \(g_{c}\) is equal to 980.6 $(mass) (cm.) / (grams force)
(sec^{2})$ and \(g\) is 978 $grams\ per\ cm$. Here \(\sigma\)
is expressed in $dynes\ per\ cm$, \(\rho_{c}\) and \(\rho_{d}\) in $grams\ per\
ml$ and \(\mu_{c}\) in $grams\ per\ cm.\ sec$.

According to Komobayasi, Gonda and Isono (1964), average life time of
water drop suspended in a vertical air stream depends on the drop
diameter. They obtained an empirical formula,
\\
\begin{equation}
    t_b=3.4\times10^6e^{-17D}
\end{equation}
\\
where \(d\) is expressed in cm and \(t_{b}\) is the life time of the
drop before it breaks in $seconds$. They also said that the number of
droplet fragments produced depends on the diameter of the drop and
obtained an empirical relation for the size distribution,
\\
\begin{equation}
    N(d_d)\delta(d_d)=6.25\times10^{-2}d_p^3e^{-7.8D} \delta d_d
\end{equation}
\\
where \(d_{d}\) and \(d_{p}\) are expressed in $cm$.

Since there is no direct measure of any non-dimensional number for the
breakup in the case of free-fall or free-rise, other factors involved
such as surface instability and the dependence on the initial shape are further considered.

\subsubsection{Surface instabilities}\label{surface-instabilities}

The breakup of particles under the action of gravity is assumed to be
due to surface instabilities. Hence considerable amount of research has
been done to study the effect of surface instability on the bubbles and
drops. Accordingly, there exists a maximum particle size above which the
particle succumbs to breakup due to surface instability. The maximum
stable bubble size is of the order of several \(cm\). Clift,
Grace and Weber (1978, ch 12.III p. 341) quote a value of
\(\sim 49\ mm\) for air bubbles in water. Grace, Wairegi and Brophy
(1978) have experimentally determined values between 45 and 90 $mm$ for
air bubbles rising in otherwise stagnant viscous media.

Different instability mechanisms have been proposed as an explanation.
The Rayleigh Taylor instability has been considered by Grace, Wairegi
and Brophy (1978) to derive a lower bound for the maximum stable
particle size as,
\\
\begin{equation}
    d_{max}=4\sqrt{\frac{\sigma}{g(\rho_l-\rho_g)}}
\end{equation}
\\
For drops (\(\lambda\  \geq 0.5\)) the data they have compiled show good
agreement with this value (see also Clift, Grace and Weber 1978, ch
12.III p 341), but for bubbles the observed maximum stable bubble sizes
are about a factor of $\sim10$ larger.

A more refined treatment of the Rayleigh-Taylor instability has been
given by Batchelor (1987) taking into account the stabilizing effect of
the liquid acceleration along the particle surface. The Kelvin-Helmholtz
instability has been considered in (Kitscha 1989) and (Wilkinson 1990).
Centrifugal forces due to the internal circulation modify the simple
treatment of instabilities above. These have been considered on a
phenomenological basis in (Levich 1962) and (Luo 1999).

\subsubsection{Dependence on initial
shape}\label{dependence-on-initial-shape}

\begin{figure}[t!]
\centering
\includegraphics[width=5in]{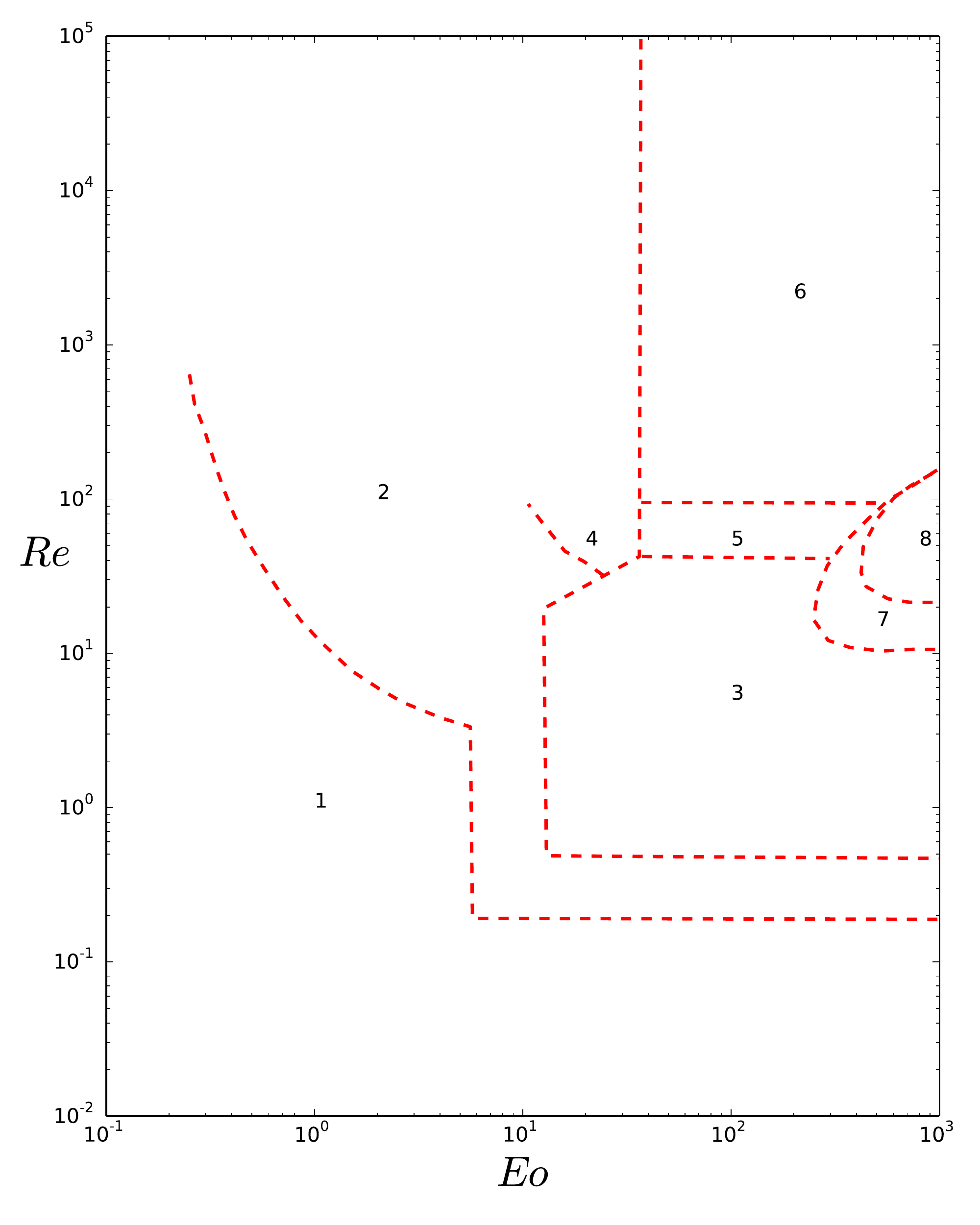}
\centering
\caption{Shapes of freely rising particle in different regimes. 1:
spherical, 2: oblate ellipsoid, 3: oblate ellipsoidal cap, 4: oblate
ellipsoidal (disk-like and wobbling), 5: spherical cap with closed,
steady wake, 6: spherical cap with open, unsteady wake, 7: skirted with
smooth, steady skirt, 8: skirted with wavy, unsteady skirt {[}Redrawn
from Figure 8 in Bhaga and Weber (1981){]}.}
\label{fig:freely_rising_particle}
\end{figure}

Final shape and state of the particles freely rising in stagnant media
have been found to depend on the initial shape of the particle. Bhaga
and Weber (1981) with the help of experiments characterized the shape of
the particle freely rising in gravity as shown in the Figure \ref{fig:freely_rising_particle}.

\begin{figure}[t!]
\centering
\includegraphics[width=4in]{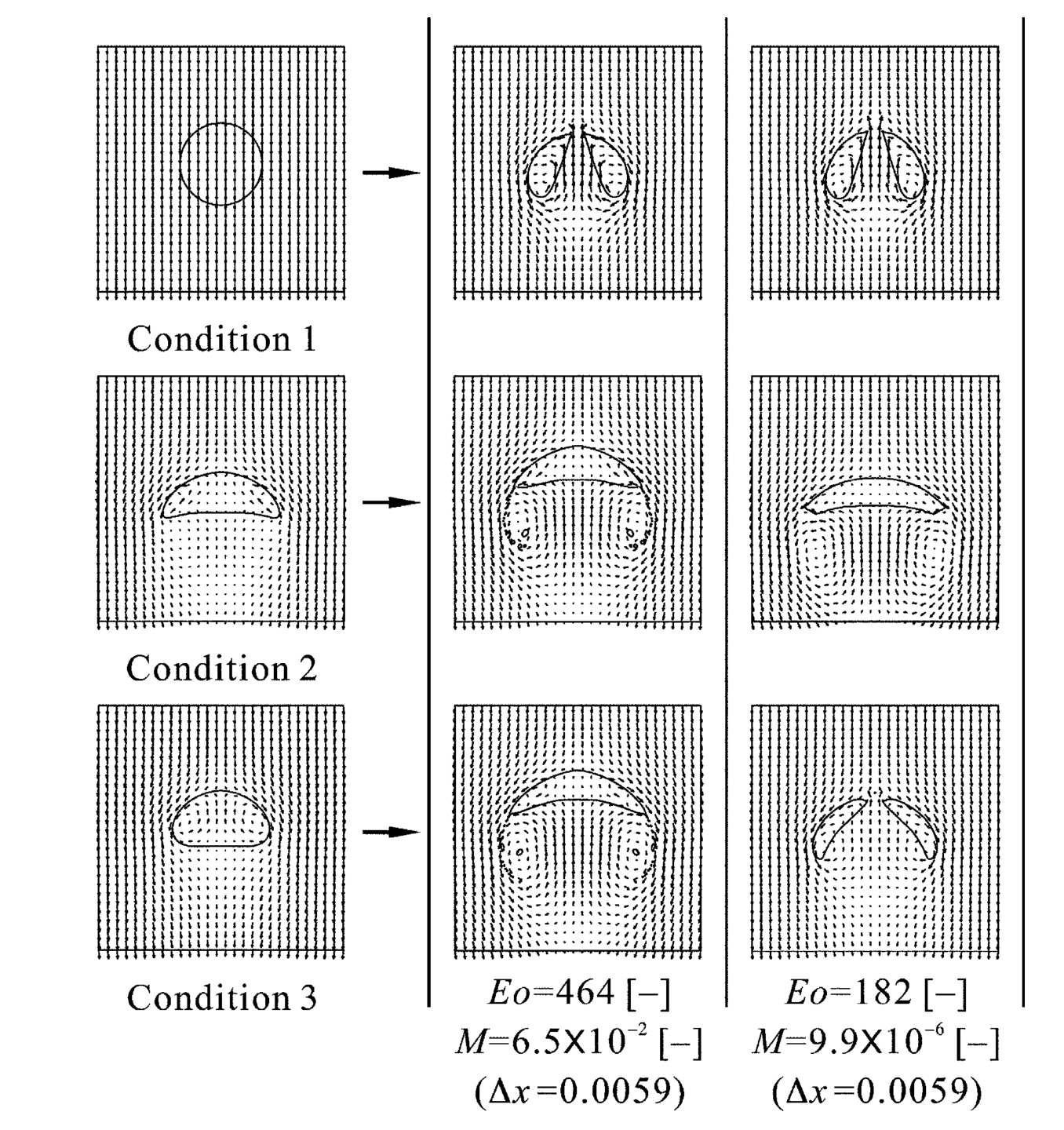}
\centering
\caption{Particle rise depending on initial shape for spherical-cap
regime. Condition 1: initially spherical particle, Condition 2 \& 3:
initially deformed particles {[}Picture taken from Ohta et al.
(2005){]}.}
\label{fig:spherical_cap}
\end{figure}

More recently, Ohta et al. (2005) studied the effect of initial shape on
the motion and final state of rising particle for which they did a
computational study using coupled level-set / volume-of fluid method.
They studied the final shape of the particle for different initial
shapes in ``spherical cap'' regime. According to their observation, the
particle breaks into a toroidal particle if it is spherical initially,
whereas an initially spherical cap particle retains its shape as shown
in the Figure \ref{fig:spherical_cap}. But the terminal velocity is found to be independent
of the initial shape and for particles from regimes other than
"spherical cap", the final shape of the particle is found to be
independent of the initial shape.

\begin{figure}[t!]
\centering
\includegraphics[width=5in]{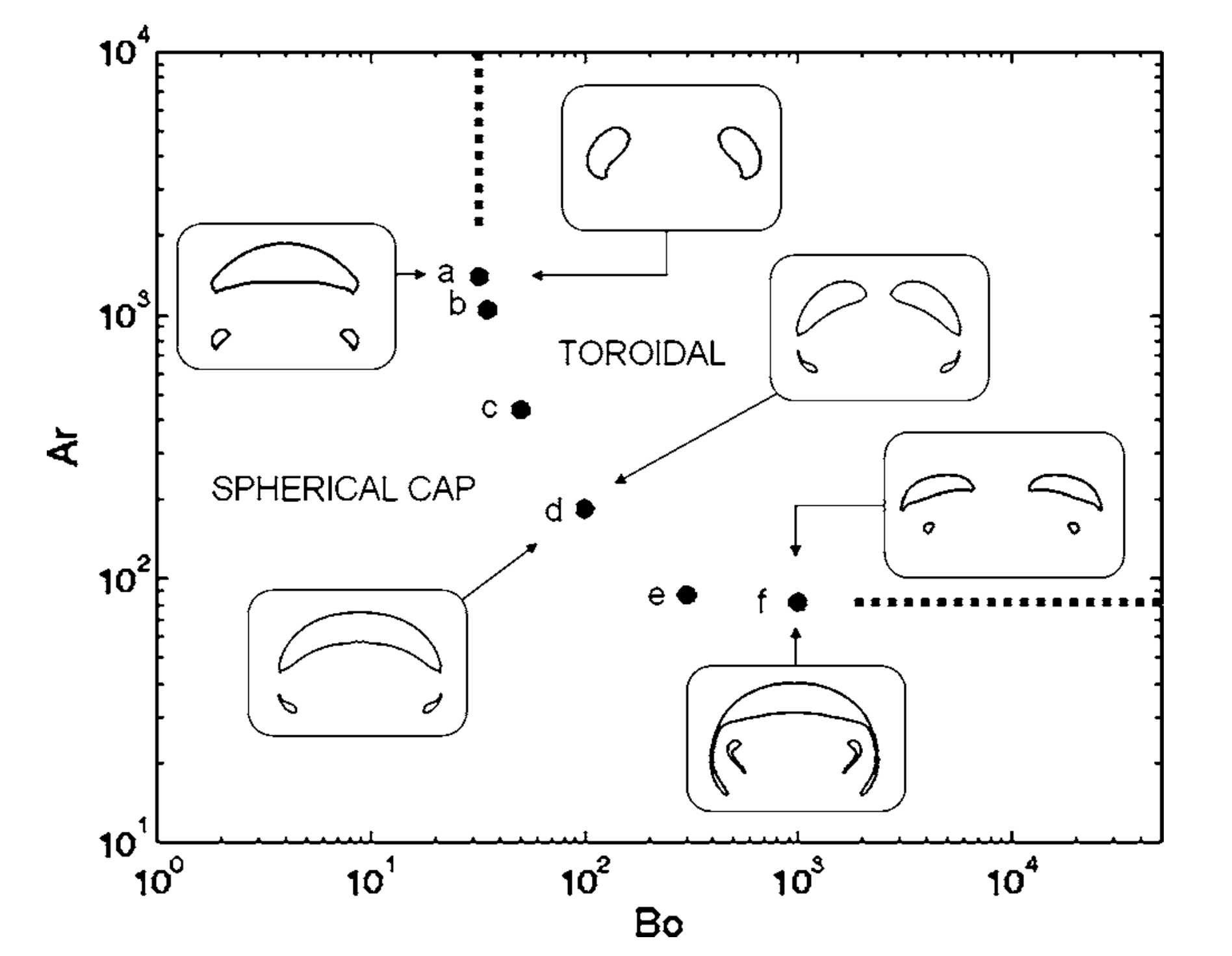}
\centering
\caption{Transition from spherical cap particles to toroidal particles.
Particle shapes are those observed just below and above the transition
line. Vertical and horizontal dashed lines indicate the location of the
transition for the inviscid regime \((Ar = \infty)\) and the
purely viscous regime \(\left(Bo = \infty \right)\),
respectively {[}Picture taken from Bonometti and Magnaudet (2006){]}.}
\label{fig:transitional_bo}
\end{figure}

Another important observation was the breakup of spherical particles in
toroidal particles, which was further numerically studied by Bonometti
and Magnaudet (2006). They produced a phase plot {[}Bond number vs
Archimedes number{]} as shown in Figure \ref{fig:transitional_bo}, indicating the transition from
the spherical cap regime to toroidal particle regime for the particles
with initially spherical shape and rising freely under the action of
gravity.

\begin{figure}[t!]
\centering
\includegraphics[width=5in]{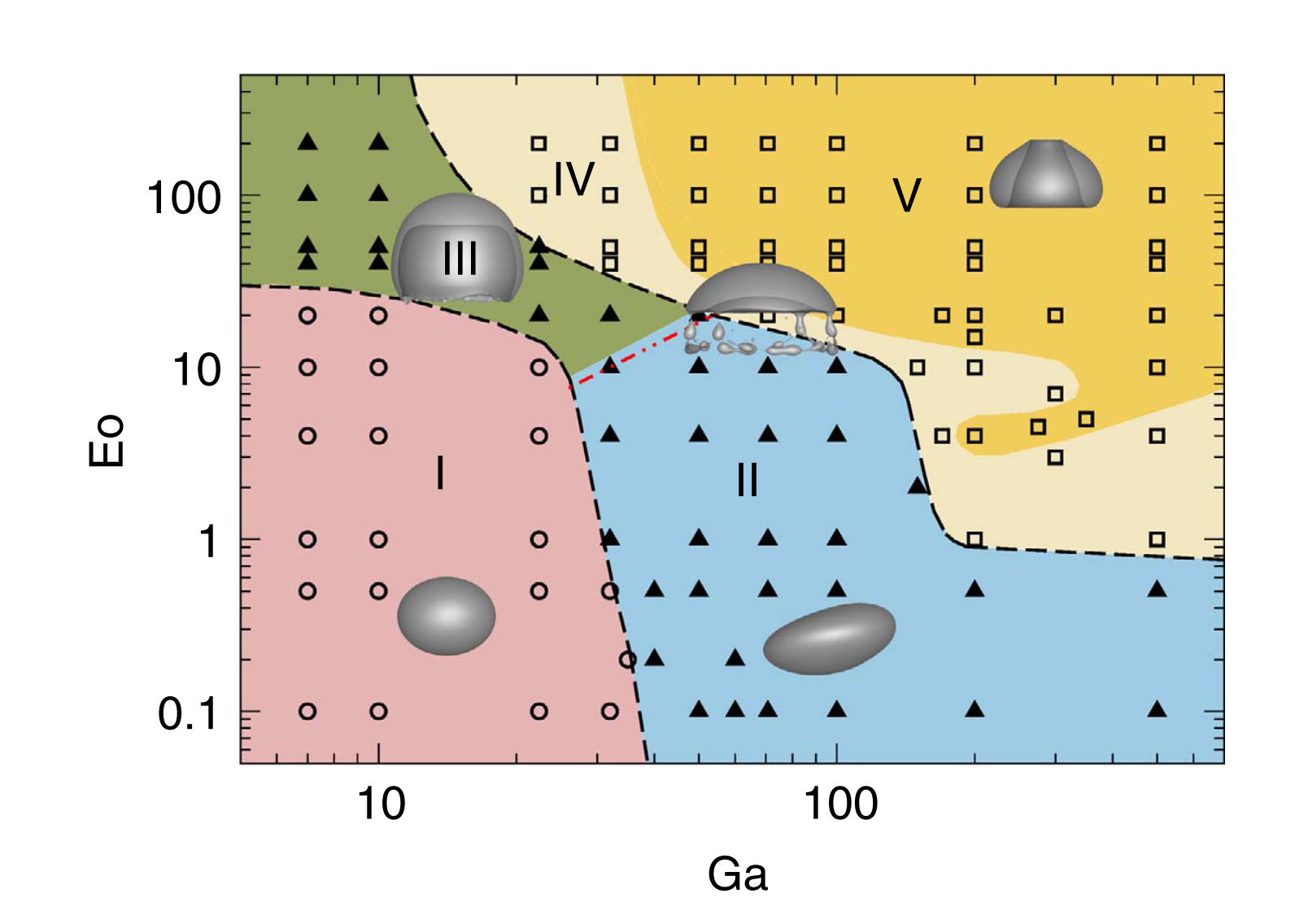}
\centering
\caption{Initially spherical particle rising freely in different
regimes. Region I: ellipsoidal in shape, axisymmetric, takes on terminal
velocity. Region II: deviates from axisymmetry, shapes change with time.
Region III: axisymmetric with thin skirt, attain terminal velocity.
Region IV: axisymmetric upto breakup, spherical cap, peripheral breakup,
axisymmetry after breakup, finally reaches terminal velocity. Region V:
doughnut or toroidal shape, small satellite particles near the boundary,
breaks into multiple particle fragments due to instability. In region
IV, for high $M$, wide skirt was formed which broke off into small
particle, for medium $M$, small particles were ejected from rim, for low
$M$, narrow skirt resulting in ellipsoid rather than cap like particle.
They used \(Eo\) with radius of the particle as the
length scale {[}Picture taken from Tripathi, Sahu and Govindarajan
(2015){]}.}
\label{fig:different_regimes}
\end{figure}

Very recently, Tripathi, Sahu and Govindarajan (2015) extended this
study to present a complete phase plot of all the regimes for an
initially spherical particle rising freely under gravity. Brief of what
was presented by them is included in Figure \ref{fig:different_regimes}.

Hence an air bubble of diameter as low as 6mm is found to breakup in
water which is contradicting to what was observed by Grace, Wairegi and
Brophy (1978). Observations on dependence of the initial shape made by
Ohta et al. (2005) explains the missing link for the stability of high
diameter bubbles observed by Grace, Wairegi and Brophy (1978), wherein
the experiments were done with initial spherical cap shape for bubbles.
Hence a bubble remains stable even above the critical diameter given by
Grace, Wairegi and Brophy (1978) if its initial shape is a spherical
cap.

%% file: turbulent.tex
\section{Breakup in complex Turbulent
flows}\label{breakup-in-complex-turbulent-flows}

Turbulent flows are the most complex and highly unsteady flows of all
types. Considering its application range, present day research mostly
deals with turbulent flows.

\subsection{Basic concepts}\label{basic-concepts}

Hinze (1955) was the first to provide a classification based on the
shape of the breakup of particles in turbulent flows into three kinds,
i.e. lenticular, cigar shape and bulgy. A universal frame work based on
static force balance was designed, according to which the deformation of
a particle is due to the external force which is counteracted by the
surface tension that tends to restore the sphericity of the particle.
Few assumptions were made to complement this theory: 1. Particles are
larger than the Kolmogorov length scale and hence the inertial effects
dominate over the viscous ones. 2. Only velocity fluctuations close to
the particle diameter are capable of causing large deformation. This
theory was independently developed by Komogorov (1949) and Hinze (1955)
and hence it is referred to as Kolmogorov-Hinze Theory.

Weber number is defined in terms of turbulent stress as,
\\
\begin{equation}
    We=\frac{\tau d}{\sigma}
\end{equation}
\\
According to the theory, further substituting the turbulent stress
\(\tau = \overline{\rho_{c}{u^{'2}}}\) and the velocity
fluctuation over a scale of the size of the particle for homogeneous
isotropic turbulence
\(\overline{u^{'2}} = C_{2}εd)^{\frac{2}{3}}\)gives
(Batchelor 1951, ch 6.5),
\\
\begin{equation}
    We=\frac{C_2 \rho_c \varepsilon^{2/3} d^{5/3}}{\sigma}
\end{equation}
\\
A critical diameter \(d_{max}\) is defined based on the critical value
of the Weber number \(We_{crit}\) at which the breakup
occurs, which was also used by Shinnar (1961):
\\
\begin{equation}
    d_{max}=We_{crit}\Big(\frac{C_2 \rho_c}{\sigma}\Big)^{-3/5} \varepsilon^{-2/5}
\end{equation}
\\
A similar theory on force balance was proposed by Levich (1962), wherein
the internal pressure of the particle is balanced with the capillary
pressure of the deformed particle. The contribution of the particle
density is included through the internal pressure force term. Hence,
\\
\begin{equation}
    We=\frac{\tau d}{\sigma} \Big(\frac{\rho_d}{\rho_c}\Big)^{1/3}
\end{equation}
\\
Risso and Fabre (1998) and Delichatsios (1975) criticized the model for
using of average value of velocity fluctuations rather than max value.
Risso and Fabre (1998), Kang and Leal (1990) and Sevik and Park (1973)
also noticed that the model neglects transient effects like resonance of
accumulation of deformation.

Walter and Blanch (1986) was one of the early works to describe the
processes involved in the particle breakup. With the help of high-speed
videos, they observed single particle breakup and suggested that the
breakup is essentially a three step process: (i) oscillation (ii)
stretching into a dumbbell shape (iii) pinching off. They gave an
expression for critical diameter, which was also confirmed by Hesketh,
Russell and Etchells (1987) and Hesketh, Etchells and Russell (1991a,
1991b):
\\
\begin{equation}
    d_{max}=1.12\frac{\sigma^{0.6}}{\Big(\frac{P}{V}\Big)^{0.4} \rho^{0.2}}\Big(\frac{\mu_c}{\mu_d}\Big)^{0.1}
\end{equation}
\\
Various correlations have been derived by other authors and are listed
in the Table \ref{tab:critical_diameter} along with the flow conditions considered.

\input{critical_diameter_table.tex}

The shape of the particles proposed by Hinze (1955) was experimentally
verified by Risso and Fabre (1998). There experiments were performed in
microgravity to minimize the bulk motion of the particle. They defined a
parameter, \(A^{*}\) - (Relative difference of the projected area with
respect to that of sphere - Equation \ref{equ:projected_area}) and performed experiments to study the
relation between the maximum projected area \(A_{\max}^{*}\ \)and the
percentage of breakup and found that there was no breakup for
\(A_{\max}^{*}\) less that 0.5, 75\% breakup for
\(A_{\max}^{*}\ \)between 0.5 and 1 and 100\% breakup for
\(A_{\max}^{*}\ \)greater than 1. Hence they proved that the particles
were significantly lengthened before they broke.
\\
\begin{equation}
    A*=\frac{4A}{\pi d^2}-1
    \label{equ:projected_area}
\end{equation}
\\
Number of fragments formed in a particle breakup was also observed by
Risso and Fabre (1998). They observed that the majority of the breakup
was binary:

\begin{enumerate}
\def\labelenumi{\arabic{enumi}.}
\item
  2 fragments (48\%).
\item
  3-10 fragments (37\%).
\item
  \textgreater{}10 fragments (15\%).
\end{enumerate}

Further breakup can be classified into 2 cases: instantaneous breakup
and resonance breakup. Though the resonance breakup was not a part of
Kolmogorov-Hinze theory initially, it was included by Risso and Fabre
(1998). Hence further detailed focus lies in these breakup processes.

\subsection{Instantaneous breakup}\label{instantaneous-breakup}

When the weber number of the flow is greater than the critical Weber
number, the sudden breakup without any significant previous deformation
is said to be an instantaneous breakup. This type of instantaneous
breakup was experimentally observed by Risso and Fabre (1998) and
numerically modelled by Higuera (2004), Revuelta, Rodriguez-Rodriguez
and Martinez-Bazan (2006), Rodriguez-Rodriguez, Gordillo and
Martinez-Bazan (2006), Galinat et al. (2006), Revuelta (2010) and
Padrino and Joseph (2011). Notable characteristics of this breakup
process are:

\begin{itemize}
\item
  This breakup is independent of the turbulence history.
\item
  The breakup might be either into two parts or many fragments.
\item
  Breakup is only due to a single eddy.
\item
  Second Eigen mode of the oscillation is largely predominant.
\item
  The breakup process is approximately axisymmetric.
\item
  Breakup is caused by the turbulent eddies of the same size as that of
  the particle.
\end{itemize}

\begin{itemize}
\item
  The flow outside the particle can be assumed to be steady since the
  characteristic turnover time \(t_{t}\) of a turbulent eddy is larger
  than the breakup time\(\text{\ t}_{b}\).
\item
  Breakup is uniaxial, and hence an initially round particle is
  stretched along a preferential direction until it breaks. This was inferred from the study
  of a biaxial breakup of a particle, which resulted in forming a disk followed
  by a torus and the breakup time was found to be larger than the uniaxial flow.
\end{itemize}

Critical weber number values have been experimentally and numerically
determined by various authors for various flow conditions. They are
listed in the Table \ref{tab:critical_we}.

\begin{table}[t!]
\centering
\caption{Critical Weber numbers for various flow conditions.}
\label{tab:critical_we}
\begin{longtable}[c]{|l|c|l|}
\hline

\textbf{Author} & \textbf{Critical Weber number} & \textbf{Flow conditions}\\ \hline

Hinze (1955) & $We_{crit} \approx 1.2$ & \begin{tabular}{c} Particles with viscosity \\ratio $\lambda  = 1$. \end{tabular} \\ \hline

Sevik and Park (1973) & $We_{crit}  \approx 2.6$ & \begin{tabular}{c}
Air bubbles in turbulent \\water jet. \end{tabular}
\\ \hline

Misksis (1981) & $We_{crit} \approx 2.76$ &
\begin{tabular} {c} Particles in steady \\axisymmetric pure straining \\flow (a simplified \\approximation to turbulent \\instantaneous breakup) \end{tabular}\\ \hline

\begin{tabular}{l} Lewis and Davidson \\ (1982) \end{tabular} & $We_{crit} \approx 4.7$
& \begin{tabular}{c} Rising gas bubbles in a \\liquid jet flowing upwards \\through a large volume of \\the same liquid. \end{tabular}\\ \hline

Ryskin and Leal (1984) & \begin{tabular}{c}
${\Big(\frac{1}{We_{crit}}\Big)}^{\frac{10}{9}} = {\Big(\frac{1}{2.76}\Big)}^{\frac{10}{9}}$ \\ 
$+ \Big(\frac{1}{0.247Re^{\frac{3}{4}}}\Big)^{\frac{10}{9}} $
\end{tabular}
& \begin{tabular} {c} Particles in steady \\axisymmetric pure straining \\flow (a simplified \\approximation to turbulent \\instantaneous breakup) \end{tabular} \\ \hline

\begin{tabular}{l} Walter and Blanch \\ (1986) \end{tabular} & $We_{crit} \approx 2.4$ &
\begin{tabular} {c} Particles in turbulent \\pipe flow.\end{tabular}\\ \hline

\begin{tabular}{l} Hesketh, Russell and \\ Etchells (1987) \end{tabular} & $We_{crit} \approx 1$ & \begin{tabular} {c} Bubbles in liquid \\in horizontal pipelines. \end{tabular} \\ \hline

Kang and Leal (1987) & \begin{tabular} {c}
$We_{crit} (Re) = 1.8 (10),$\\ $4.2 (100), 5.4 (\infty)$
\end{tabular}
& \\ \hline

Risso and Fabre (1998) & $We_{crit}  = 2.7\sim 2.8$ & \begin{tabular} {c} Particles in microgravity \\condition.
\end{tabular} \\ \hline

\begin{tabular}{l} Revuelta, Rodriguez-\\ Rodriguez and \\ Martinez-Bazan (2006) \end{tabular} & $We_{crit} = 2.2$ for $Re  \geq 20$ & \begin{tabular} {c} Numerically calculated \\using USF (Uniaxial \\Straining Force) model. 
\end{tabular}\\ \hline

\end{longtable}
\end{table}

\paragraph{Uniaxial Straining Force (USF) Model}\label{USF}

Following the idea of
static force balance, Martinez-Bazan, Montanes and Lasheras (1999)
developed an empirical model for breakup frequency at high Reynolds
number and Weber number, based on the difference between the turbulent
stress and the surface stress.

for \(Re \rightarrow \infty\)
\\
\begin{equation}
    t_b\approx\Big(1-\frac{We_{crit}}{We}\Big)^{-\frac{1}{2}}
\end{equation}
\\
\\
\begin{equation}
    g*\approx\Big(1-\frac{We_{crit}}{We}\Big)^{\frac{1}{2}}
\end{equation}
\\
Where g* is equal to dimensionless breakup frequency. This was
validated recently by Padrino and Joseph (2011), who used a Boundary
element method to perform a viscous irrotational analysis. This model has been widely accepted and is being used as the breakup model for turbulent flows in most of the commercial and open source CFD solvers.

Further, Revuelta, Rodriguez-Rodriguez and Martinez-Bazan (2006)
extended the model for finite $Re$ values using a level-set
numerical scheme. They analyzed the influence of the steady,
axisymmetric pure straining flow on a particle and found that the
critical value of $We$ is 2.2 for \(Re\  \geq \ 20\) and that
the breakup time was significantly different for finite Reynolds numbers
as shown in the Figure 18. Finally, they included the effect of Reynolds
number in the expression of the breakup frequency.
\\
\begin{equation}
    t_b\approx\Big(1+\frac{15.5}{Re}\Big)\Big(1-\frac{We_{crit}}{We}\Big)^{-\frac{1}{2}}
    \label{equ:breakup_time}
\end{equation}
\\
\begin{figure}[t!]
\centering
\includegraphics[width=5in]{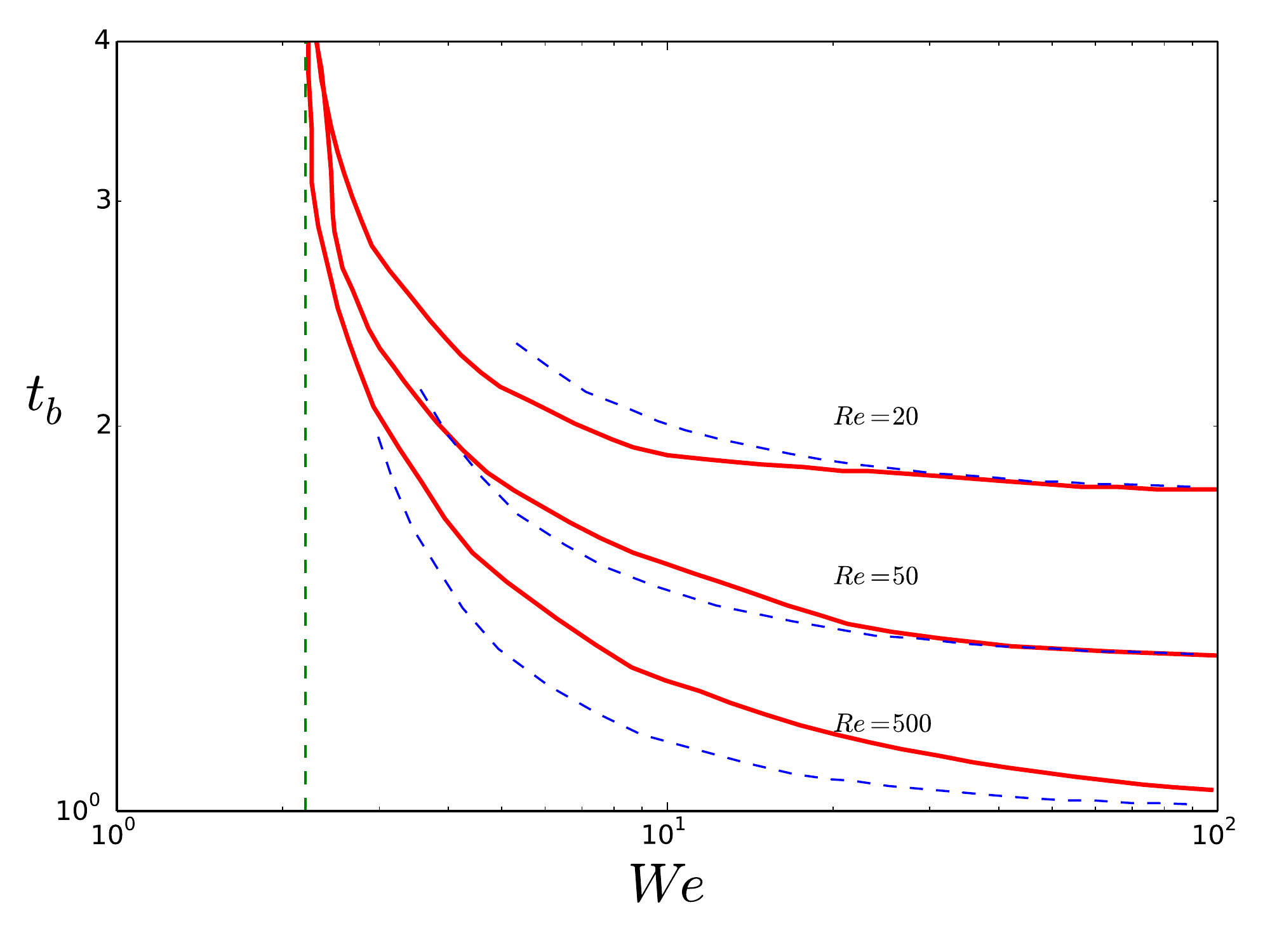}
\centering
\caption{Dependence of breakup time \(t_{b}\) of a
particle on the weber number \(We\) and Reynolds number
\(Re\) for supercritical Weber number. Solid lines:
results obtained by Revuelta, Rodriguez-Rodriguez and Martinez-Bazan
(2006). Dashed blue lines: approximated relation in Eq. (30). Dashed
green line: critical Weber number {[}Redrawn from Figure 5 in Revuelta,
Rodriguez-Rodriguez and Martinez-Bazan (2006){]}.}
\label{fig:breakup_time}
\end{figure}

\paragraph{Effective size range:}\label{effective-size-range}

Though it was presumed that the breakup is caused by turbulent eddies of same size, Higuera (2004) went on to
numerically study the collision of an axisymmetric stationary particle
with a vortex ring of varying sizes to further explore the behavior of
the breakup. Vortex ring was assumed to be an approximation for a
turbulent eddy, though they knew that it is far from realistic
collisions, they tried to make a qualitative study. They found that,

\begin{itemize}
\item
  For\(\ R_{\text{vp}}\  < \ 0.75\), breakup was into 2 daughter
  particles of approximately same size.
\item
  For\(\ 0.75\  < \ R_{\text{vp}}\  < \ 1.1\), particle tears off.
\item
  For\(\ R_{\text{vp}}\  > \ 1.1\), a hole is punched at the symmetry
  axis, which transforms the particle into an elongated torus.
\end{itemize}

where \(R_{\text{vp}}\) is the ratio of the radius of the vortex ring to
that of the particle. Finally correlating the translatory velocity with
the velocity scale at inertial range of turbulent flow, they arrived at
the expression,
\\
\begin{equation}
    We=We_bR_{vp}^{\frac{2}{3}}
\end{equation}
\\
where
\(\text{We}_{b}\sim\frac{\rho\varepsilon^{\frac{2}{3}}a^{\frac{5}{3}}}{\sigma}\)
is the Weber number for vortices of radius equal to that of the
particle. Intersection of the plot of this weber number with that of the
critical weber number gives a range for the radius of the vortex which
is effective in breaking the particle. Rodriguez-Rodriguez, Gordillo and
Martinez-Bazan (2006) criticized their study by stating that the breakup
time strongly depends on the initial position of the vortex core and
that they failed to supply significant information about real dependence
of the breakup time on the parameters of the problem.

Further recently, Revuelta (2010) performed full 3D simulations of the
same and presented a better plot of particle breakup patterns for different size ratios (Figure 19), which the USF model had failed to account for.

\begin{figure}[H]
\centering
\includegraphics[width=5in]{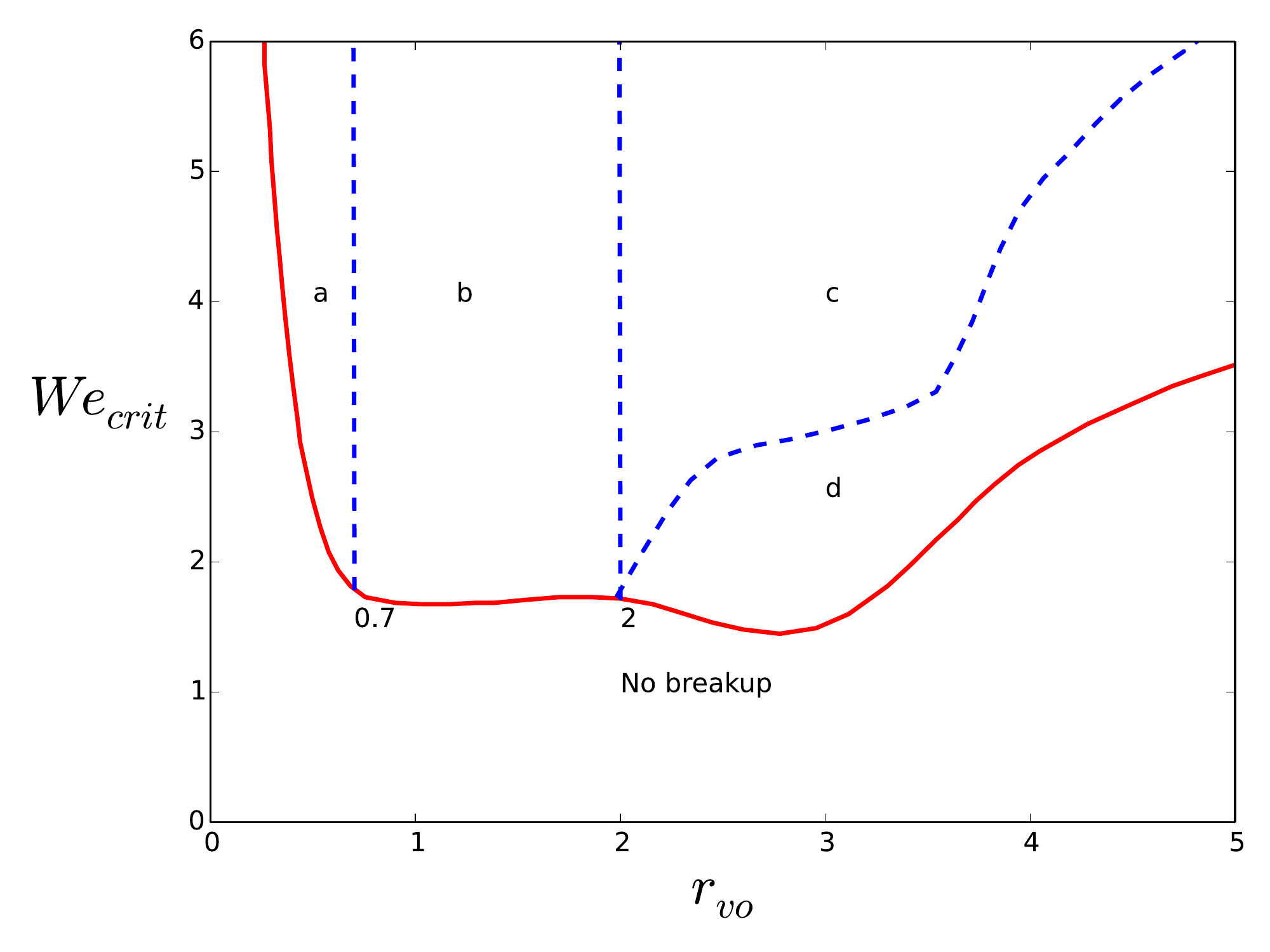}
\centering
\caption{Critical Weber number
\(We_{crit}\) vs the initial vortex-ring
radius \(r_{vo}\) for \(Re = 1000\),
\(\overline{\delta}=\frac{\delta}{a}= 0.05\). Region a: no binary breakup, no cigar formation, only satellite
particles. Region b: lenticular deformation, multiple fragment breakup.
Region c: lenticular formation, multiple similar size fragments,
inverted U shape probability density function. Region d: cigar
formation, U shape probability density function, binary non-similar size
fragments. No breakup below the line {[}Redrawn from Figure 1 in
Revuelta (2010){]}.}
\label{fig:critical_we}
\end{figure}

Critical Weber number obtained was in the same order as that of USF
model. Hence this explains by far most of the phenomenon observed in
experiments, though a non-aligned collision of a vortex ring and
particle was not completely studied.

\subsection{Resonance breakup}\label{resonance-breakup}

For the flows with $We>We_{crit}$, the
parent particle breaks into two or more daughter particles and the
relation derived for $t_b$ (Equation \ref{equ:breakup_time}) can be used to determine the approximate breakup time. But for sub-critical
Weber numbers, individual turbulent eddies are not capable enough to break the
interface, and the particle oscillates with an increasing frequency and
decreasing amplitude as We decreases. However, the particle could still
break due to previous deformation or fluctuating strength of the outer
flow field. Hence the breakup depends on \(t_{r}\) (residence time),
\(t_{\text{rp}}\ \)(particle response time), damping rate and turbulent
time scales. This kind of breakup is referred to as resonance breakup.

Sevik and Park (1973) were the first to postulate resonance mechanism
between the particle dynamics and the turbulent fluctuations. Later,
Risso and Fabre (1998) coupled Rayleigh-Lamb theory (Lamb 1932 -- valid
up to breakup point) of particle oscillation with the Kolmogorov-Hinze
theory of turbulent breakup to explain interface deformation for a
particle in a homogeneous turbulent field. Turbulent excitation was
modeled with the same assumption as Kolmogorov-Hinze theory. Finally,
they extended the Komogorov-Hinze theory for resonance breakup:

\begin{enumerate}
\def\labelenumi{\alph{enumi}.}
\item
  (Weak turbulence) \(We C_{d}(t\rightarrow \infty)\)
  \textless{} Critical value:

  Individual eddies are not capable of causing breakup and accumulation
  of energy is not sufficient. So breakup is impossible.
\item
  (Moderate turbulence) \(We C_{d}(t \rightarrow \infty)\)
  \textgreater{} Critical value and
  \(\text{We}C_{d}(t\  \rightarrow 0)\) \textless{} Critical value:

  Individual eddies are not capable of causing breakup, although a
  succession of turbulent eddies may be. Breakup is controlled by
  resonance like mechanism.
\item
  (Intense turbulence) \(We C_{d}(t \rightarrow 0)\)
  \textgreater{} Critical value:

  Individual eddies are capable of causing sudden breakup of a
  previously non-deformed particle. Breakup agrees with force balance
  interpretation.
\end{enumerate}

where \(C_{d}\) is mean efficiency coefficient, \(C_{d}\) depends on
particle diameter and residence time of the particle.

Galinat et al. (2007) tested Risso and Fabre (1998)'s theory to the case
of particle travelling through an inhomogeneous turbulent field and to
that of concentrated particle dispersions and proved validity of the
theory for a more general case. Revuelta, Rodriguez-Rodriguez and
Martinez-Bazan (2006) numerically simulated the resonance breakup at
subcritical Weber number and observed that the particle deformation
increases with the amplitude \(\zeta\) of the external excitation. Residence
time of particle (i.e. the time either particle breaks or manages to escape from the external flow) within the pure straining flow
was found to decrease with the increase in \(\zeta\). The amplitude required for
breakup \(\zeta_{b}\) increases with the reduction of $Re$ due to attenuation
effect of the viscosity and \(\zeta_{b}\) decreases as $We$ increases
towards critical value.

Further, Revuelta, Rodriguez-Rodriguez and Martinez-Bazan (2008)
modelled turbulent breakup of particles at sub-critical Weber number
using a 2D model of particle subjected to a pulsating USF with strain
direction varying randomly for every pulse. They found that the relevant
parameter that determines whether the particle breaks or not at a given
time is the instantaneous surface energy which is directly correlated to
the interfacial length, independently of the intensity of the outer flow
or the detailed history of the deformation. When the frequency of the
eddies train is similar to the oscillation frequency of the particle
immersed in a steady USF whose Weber number is equal to the effective
Weber number of the pulsating USF, they observed maximum breakup and
breakup rate which is due to resonance. At sub-critical Weber number,
they also observed similar characteristics in all breakup patterns:
initially the particle adopts a cigar shape before taking the form of a
dumb bell and finally breaks into fragments of approximately the same
size. They also studied the effect of difference of angle between two
consecutive strain directions: for an angle of around 90 deg, particle
increases its level of deformation breaking up at values of critical
surface energy whereas for angle of around 0 deg, no breakup is seen.

\subsection{Breakup under the action of
gravity:}\label{breakup-under-the-action-of-gravity}

Breakup of particles under the action of gravity in stagnant medium has
already been covered in the Section \ref{breakup-under-the-action-of-gravity-in-stagnant-media} but for most of the present day
practical (industrial) purposes, flow in a bubble column is generally
governed both by the buoyancy and turbulence.

Until recently most of the studies concentrated on studying particle in
a turbulent jet flow and pipe flows and accordingly modeling of these
flows dealt with particle breakup assuming microgravity conditions. But
Ravelet, Colin and Risso (2011), was the first to experimentally study
the effect of buoyancy on the breakup of particles in turbulent flow.
Assuming the gravity conditions for a flow in bubble column, buoyancy
significantly affects the breakup mechanism due to relative slip between
the particle and the external fluid.

According to the assumptions of the USF model, particle is moving at
local mean fluid velocity (particle Reynolds number is zero) and the
characteristic size of the particle lies within the inertial subrange of
the turbulent energy spectrum. Hence the Reynolds number based on outer
fluid properties of the flow surrounding the particle is sufficiently
large to neglect the viscous effects. But this situation could be
completely different for a particle rising in turbulent flow where the
Reynolds number based on outer fluid properties is very low to neglect
viscous effects and the particle Reynolds number is not zero but very
high due to significant slip.

Characteristics of the flow and breakup in a turbulent bubble column
are:

\begin{itemize}
\item
  Particle deformation dynamics is radically different from that
  observed in the absence of significant sliding motion due to buoyancy. Large deformations that lead to breakup are not axisymmetric.
\item
  No trace of periodic oscillation at the eigen frequency \(f_{2}\),
  contrary to what is observed without buoyancy.
\item
  Value of critical deformation leading to breakup is in agreement with
  results without buoyancy for particle in homogenous turbulence.
\item
  The timescale of decay of shape oscillations is of order of their
  natural frequency $f_2$, hence there is no stochastic
  resonance and breakup results only from the interaction with a single
  turbulent eddy.
\item
  The magnitude of order of the instantaneous weber number was observed
  to be around 10.
\end{itemize}

Scaling of the particle breakup \(t_{b}\) was observed to be different
depending on the interaction time \(t_{i}\) of the particle with a
turbulent eddy:

\begin{itemize}
\item
  If \(t_{i}\) is short compared to \(\frac{1}{f_{2}}\), \(t_{b}\) is
  proportional to \(\frac{1}{f_{2}}\).
\item
  If \(t_{i}\) is large compared to \(\frac{1}{f_{2}}\), the forcing
  flow can be considered steady and USF model might be reasonable.
\end{itemize}

Further in turbulent flows, many other classification of flow types are
observed, studied and reviewed. Experimental studies, theoretical
breakup models and their derived correlations for drops and bubbles in
turbulent pipe flows and in stirred vessels are few such examples. But
these are considered beyond the scope of this article, since they are
too problem specific and do not in general explain the actual physics of
breakup process.

%% file: critical_diameter_table.tex
\begin{table}[H]
\centering
\caption{Correlations for critical diameter derived for various flow
conditions.}
\label{tab:critical_diameter}
\begin{longtable}[c]{|l|c|l|}
\hline

\textbf{Author} & \textbf{Max diameter} & \textbf{Flow conditions} \\
\hline
\endfirsthead

\textbf{Author} & \textbf{Max diameter} & \textbf{Flow conditions} \\
\hline
\endhead

\begin{tabular}[c]{@{}l@{}}Hinze (1955), \\ Shinnar (1961)\end{tabular}                                                                 & \begin{tabular}{c}
$d_{max} = We_{crit}(\frac{C_{3}\rho_{c}}{\sigma})^{-\frac{ 3}{5}}\varepsilon^{-\frac{2}{5}}$, \\ where $C_{3} = \text{contant}$
\end{tabular} & \begin{tabular}[c]{@{}l@{}}Particles with\\ viscosity ratio\\ $\lambda=1$\end{tabular}                                                             \\ \hline

\begin{tabular}[c]{@{}l@{}}Sleicher \\ (1962)\end{tabular}                                                                              &  \begin{tabular}{c}
$\frac{d_{max}\rho_{c}V^{2}}{\sigma}\sqrt{\frac{\mu_{c}V}{\sigma}} = C_{4}\left\lbrack 1 + 0.7\left( \frac{\mu_{d}V}{\sigma} \right)^{0.7} \right\rbrack$ \\
$C_{4}$ = 38  with  1.5  inch  I.D of pipe.
\end{tabular}  & \begin{tabular}[c]{@{}l@{}}Drops of two\\ immiscible liquids\\ in turbulent\\ pipe flow\end{tabular}                                                \\ \hline

Paul (1965)                                          &
\begin{tabular}{c}
$\frac{d_{max}\rho_{c}V^{2}}{\sigma}\sqrt{\frac{\mu_{c}V}{\sigma}} = C_{5}\left\lbrack 1 + 0.7\left( \frac{\mu_{d}V}{\sigma} \right)^{0.7} \right\rbrack$ \\
$C_{5}$ = 43  with  1.5  inch  I.D of pipe.\\
$C_5$ is dependent on the pipe diameter.
\end{tabular}
& \begin{tabular}[c]{@{}l@{}}Drops of two\\ immiscible liquids\\in turbulent\\ pipe flow\end{tabular} \\ \hline

Wallis (1969)   &  $d_{max} = 0.752\sigma^{0.6}\varepsilon^{- 0.4}\rho^{- 0.2}$ & Particle in liquid   \\ \hline

\begin{tabular}[c]{@{}l@{}}Kubie and \\ Gardner \\ (1977)\end{tabular}  &          \begin{tabular}{c}
$\left\lbrack \frac{d_{max}\rho_{c}\mathbf{u}^{2}}{\sigma} \right\rbrack\left\lbrack \frac{fd_{\max}}{D} \right\rbrack^{\frac{2}{3}} = 0.369$\\
$f = 0.076Re^{- 0.25}$ for straight pipe\\
$f = 0.076Re^{- 0.25} + 0.00725\left( \frac{D}{D_{H}} \right)^{0.5}$\\
for helix pipe
\end{tabular} 
& \begin{tabular}[c]{@{}l@{}}Drops of \\ liquid-liquid \\ system in pipe flow\end{tabular}                                                            \\ \hline

Arai et al. (1977)   &
\begin{tabular}{c}
\(\frac{d_{\max}}{D_{i}} = C_{6}.\left( \frac{\rho_{c}n_{r}{D_{i}}^{2}}{\mu_{d}} \right)^{- 0.75}\)\\
when dispersed phase viscosity $\mu_{d}$\\
is not negligible considerable
\end{tabular}
& \begin{tabular}[c]{@{}l@{}}Particles in \\ mixing vessel\end{tabular} \\ \hline

\begin{tabular}[c]{@{}l@{}}Lewis and \\ Davidson (1982)\end{tabular} &
\begin{tabular}{c}
$d_{\max} = 1.67\sigma^{0.6}\varepsilon^{- 0.4}\rho^{- 0.2}$\\
Industrial bubble columns \\
$(\varepsilon \approx 1kW/m^{3})$
- $d_{\max} = 5.5mm$, \\
Laboratory stirred tank \\
$(\varepsilon \approx 10kW/m^{3})$
- $d_{\max} = 2.2mm$
\end{tabular}
& \begin{tabular}[c]{@{}l@{}}Rising gas bubbles \\ in a liquid jet \\ flowing upwards \\ through a large \\ volume of the \\ same liquid\end{tabular} \\ \hline

Davies (1985) & 
$d_{\max} = C_{7}*\left( \sigma + \frac{\mu_{d}u^{'}}{4} \right)^{0.6}\rho_{c}^{- 0.6}\ \varepsilon^{- 0.4}$ &
\begin{tabular}[c]{@{}l@{}}Particle in liquid \\ emulsions\end{tabular} \\ \hline

\begin{tabular}[c]{@{}l@{}}Walter and \\ Blanch (1986)\end{tabular}&
$d_{max} = 1.12\frac{\sigma^{0.6}}{\left( \frac{P}{V} \right)^{0.4}\rho^{0.2}}\ \left(\frac{\mu_{c}}{\mu_{d}} \right)^{0.1}$
& \begin{tabular}[c]{@{}l@{}}Particles in \\ turbulent pipe \\ flow\end{tabular}
\\ \hline

\begin{tabular}[c]{@{}l@{}}Calabrese, \\ Wang and \\ Bryner (1986), \\ Sathyagal, \\ Ramkrishna \\ and Narsimhan \\ (1996)\end{tabular} &
\begin{tabular}{c}
$\frac{d_{32}}{d_{o32}} = 0.053We^{- 0.6}\left( 1 + 4.42Oh \right)^{3/5}$\\
where $Oh = \frac{\sqrt{\frac{\rho_c}{\rho_d}}\mu_d\varepsilon^{-\frac{1}{3}}}{\sigma}d_{32}^{\frac{1}{3}}$\\
as $Oh \rightarrow 0$  $\frac{d_{32}}{d_{o32}} = 0.053We^{- 0.6}$\\
as $Oh \rightarrow \infty$  $\frac{d_{32}}{d_{o32}} = 0.13\left( \frac{\rho_{c}}{\rho_{d}} \right)^{\frac{3}{8}}\left( \frac{\mu_{d}}{\mu_{c}} \right)^{\frac{3}{4}}Re^{-\frac{3}{4}}$\\
and $\frac{d_{32}}{d_{max}}$ = 0.48 to 0.6\\
depending on the value of the $\mu_{d}$ 
\end{tabular}
& \begin{tabular}[c]{@{}l@{}}Particles in \\ turbulent \\ stirred tanks\end{tabular}\\ \hline

Hesketh (1987)&
\begin{tabular}{c}
$d_{max} = \left(\frac{We_{\text{crit}}}{2} \right)^{0.6}\left\lbrack \sigma^{0.6}/\ {(\rho}_{c}^{2}\rho_{d} \right)^{0.2}\ \rbrack\varepsilon^{- 0.4}$\\
$with\ We_{crit}$ = 1.1
\end{tabular}
& \begin{tabular}[c]{@{}l@{}}Bubbles in \\ liquids in \\ horizontal \\ pipelines\end{tabular} \\ \bottomrule

\end{longtable}
\end{table}

%% file: references.tex
\section*{Nomenclature}

\textit{$a$}{\ \  Radius of the initially spherical particle / unperturbed jet ($m$)} \\\\
\textit{$Ar$}{\ \  Archimedes number} \\\\
\textit{$A^{*}$}{\ \  Relative difference of the projected area of the particle with respect to that of sphere ($m^2$)} \\\\
\textit{$A_{\max}^{*}$}{\ \  Maximum relative difference of the projected area of the particle with respect to that of sphere ($m^2$)} \\\\
\textit{${Bo}$}{\ \  Bond number}
\textit{$B$}{\ \  Half-breadth of the particle ($m$)} \\\\
\textit{$Ca$}{\ \  Capillary number} \\\\
\textit{$C_{d}$}{\ \  Mean efficiency coefficient} \\\\
\textit{$d$}{\ \  Particle diameter ($m$)} \\\\
\textit{$d_{d}$}{\ \  Daughter particle diameter ($m$)} \\\\
\textit{$d_{p}$}{\ \  Parent particle diameter ($m$)} \\\\
\textit{$d_{\max}$}{\ \  Maximum stable particle diameter ($m$)} \\\\
\textit{$d_{o32}$}{\ \  Sauter mean diameter of the inviscid dispersed phase ($m$)} \\\\
\textit{$d_{32}$}{\ \  Sauter mean diameter of the dispersed phase ($m$)} \\\\
\textit{$D$}{\ \  Cross section diameter of the jet ($m$)} \\\\
\textit{${De}$}{\ \  Deformation of the particle} \\\\
\textit{$D_{H}$}{\ \  Helix diameter ($m$)} \\\\
\textit{$D_{i}$}{\ \  Impeller diameter in stirred tanks ($m$)} \\\\
\textit{$D_{t}$}{\ \  Pipe / tube diameter ($m$)} \\\\
\textit{$Eo$}{\ \  Eötvös number} \\\\
\textit{$f_{2}$}{\ \  Eigen frequency of the second mode ($s^{- 1}$)} \\\\
\textit{$g$}{\ \  Acceleration due to gravity (${m}s^{- 1}$)} \\\\
\textit{$g^{*}$}{\ \  Dimensionless breakup frequency} \\\\
\textit{$G$}{\ \  Shear rate (strength of shear flow) ($s^{- 1}$)} \\\\
\textit{${Ga}$}{\ \  Galilei number} \\\\
\textit{$g_{c}$}{\ \  A conversion factor} \\\\
\textit{$k$}{\ \  Wavenumber ($m^{- 1}$)} \\\\
\textit{$k_{c}$}{\ \  Cut-off wavenumber ($m^{- 1}$)} \\\\
\textit{$l_{\frac{1}{\ \  2}}$}{\ \  Initial half length of the particle ($m$)} \\\\
\textit{$L$}{\ \  Half-length of the particle ($m$)} \\\\
\textit{$M$}{\ \  Morton number} \\\\
\textit{$n$}{\ \  Growth rate of disturbance ()} \\\\
\textit{$n_{r}$}{\ \  Impeller speed ($s^{- 1}$)} \\\\
\textit{${Oh}$}{\ \  Ohnesorge number} \\\\
\textit{$P$}{\ \  Pressure (${Pa}$)} \\\\
\textit{$r$}{\ \  Radial coordinate ($m$)} \\\\
\textit{$r_{{vo}}$}{\ \  Initial vortex ring radius ($m$)} \\\\
\textit{${Re}$}{\ \  Reynolds number} \\\\
\textit{$R_{v}$}{\ \  Vortex ring radius ($m$)} \\\\
\textit{$R_{{vp}}$}{\ \  Ratio of vortex ring radius to the particle radius} \\\\
\textit{$S_{d}$}{\ \  A dimensional physical group}
\textit{$t$}{\ \  Time ($s$)} \\\\
\textit{$t_{b}$}{\ \  Breakup time or lifetime of the particle ($s$)} \\\\
\textit{$t_{i}$}{\ \  Interaction time of a particle with a turbulent eddy ($s$)} \\\\
\textit{$t_{r}$}{\ \  Particle residence time ($s$)} \\\\
\textit{$t_{t}$}{\ \  Characteristic turnover time of a turbulent eddy ($s$)} \\\\
\textit{$t_{r}$}{\ \  Particle response time ($s$)} \\\\
\textit{$u$}{\ \  mean velocity (${m}s^{- 1}$)} \\\\
\textit{$U$}{\ \  Generic velocity scale (${m}s^{- 1}$)} \\\\
\textit{$u'$}{\ \  fluctuation velocity (${m}s^{- 1}$)} \\\\
\textit{$U_M$}{\ \  Terminal velocity of the drop at max diameter (${m}s^{- 1}$)} \\\\
\textit{$V$}{\ \  Volume ($m^3$)} \\\\
\textit{$We$}{\ \  Weber number} \\\\
\textit{$x$}{\ \  Coordinate along main flow direction ($m$)} \\\\
\textit{$y$}{\ \  Coordinate normal to the wall ($m$)} \\\\
\textit{$z$}{\ \  Third spatial coordinate ($m$)} \\\\
\textit{$\alpha$}{\ \  Flow type parameter} \\\\
\textit{$\Gamma_{\infty}$}{\ \  Upper bound to surfactant concentration (${mol}m^{- 2}$)} \\\\
\textit{$\delta$}{\ \  Vortex core radius ($m$)} \\\\
\textit{$\overline{\delta}$}{\ \  Ratio of vortex core radius to the initial particle radius} \\\\
\textit{$\varepsilon$}{\ \  Turbulent dissipation ($m^{2}s^{- 3}$)} \\\\
\textit{$\zeta$}{\ \  External excitation amplitude ($m$)} \\\\
\textit{$\zeta_{0}$}{\ \  Initial perturbation amplitude ($m$)} \\\\
\textit{$\zeta_{b}$}{\ \  External excitation amplitude at breakup ($m$)} \\\\
\textit{$\lambda$}{\ \  Viscosity ratio} \\\\
\textit{$\mu$}{\ \  Dynamic viscosity} \\\\
\textit{$\nu$}{\ \  Kinematic viscosity ($m^{2}s^{- 1}$)} \\\\
\textit{$\rho$}{\ \  Density (${kg}m^{- 3}$)} \\\\
\textit{$\sigma$}{\ \  Surface tension (${N}m^{- 1}$)} \\\\
\textit{$\tau$}{\ \  Turbulent stress (${N}m^{- 2}$)} \\\\
\textit{$\omega$}{\ \  Growth rate of disturbance} \\\\

\section*{Subscripts}

\textit{$crit$}{\ \  critical value}\\\\
\textit{$c$}{\ \  continuous phase}\\\\
\textit{$d$}{\ \  dispersed phase}\\\\

\section*{REFERENCES}\label{references}

Acrivos, A \& Lo, T, S 1978, `Deformation and breakup of a slender drop
in an extensional flow', J. Fluid Mech., Vol. 86, pp. 641-672.

Acrivos, A 1983, `The breakup of small drops and bubbles in shear
flows', Ann. N. Y. Acad. Sci., Vol. 404, pp. 1-11.

Arai, K, Konno, M, Matunga, Y \& Saito, S, 1977, `Effect of
dispersed-phase viscosity on the maximum stable drop size for breakup in
turbulent flow', J. Chem. Eng. Japan, Vol. 10, pp. 325-330.

Ashgriz, N \& Mashayek, F 1995, `Temporal analysis of capillary jet
breakup', J. Fluid Mech., Vol. 291, pp. 163-190.

Ashgriz, N (ed.) 2011, `Handbook of atomization and sprays', Springer.

Barthes-Biesel, D 1972, PhD thesis, Stanford University.

Barthes-Biesel, D \& Acrivos, A 1973, `Deformation and burst of a liquid
droplet freely suspended in a linear shear field', J. Fluid Mech., Vol.
61, part 1, pp. 1-21.

Batchelor, G, K 1959, `The theory of homogeneous turbulence', Cambridge
University Press.

Batchelor, G, K 1987, `The stability of a large gas bubble rising
through liquid', J. Fluid Mech., Vol. 184, pp. 399-422.

Bentley, B, J \& Leal, L, G 1986, `An experimental investigation of drop
deformation and breakup in steady two-dimensional linear flows', J.
Fluid Mech., Vol. 61, pp. 1-21.

Bhaga, D \& Weber, M, E 1981, `Bubbles in viscous liquids: shapes, wakes
and velocities', J. Fluid Mech. Vol. 105, pp. 61-85.

Bonometti, T \& Magnaudet, J 2006, `Transition from spherical cap to
toroidal bubbles', Phys. Fluids, Vol. 18, pp. 1-12.

Bousfield, D, W, Keunings, R, Marrucci, G \& Denn, M, M 1986, `Nonlinear
analysis of the surface tension driven breakup of viscoelastic
filaments', J. Non-Newtonian Fluid Mech., Vol. 21, pp. 79-97.

Bhaga, D \& Weber, M, E 1981, `Bubbles in viscous liquids: shapes, wakes
and velocities', J. Fluid Mech., Vol. 105, pp. 61-85.

Buckmaster, J, D 1972, `Pointer bubbles in slow viscous flow', J. Fluid
Mech., Vol. 55, pp.385-400.

Buckmaster, J, D 1973, `The bursting of pointer drops in slow slow
viscous flow', J. Appl. Mech., Vol. 40, pp. 18-24.

Calabrese, R, V, Wang, C, Y \& Bryner, N, P 1986, `Drop breakup in
turbulent stirred-tank contactors. Part 1: Effect of dispersed-phase
viscosity', AIChE J., Vol. 32, pp. 677-681.

Cao, X, K, Sun, Z, G, Li, W, F, Liu, H, F, Yu, Z, H 2007, `A new breakup
regime for liquid drops identified in a continuous and uniform air jet
flow', Phys. Fluids, Vol. 19(5), pp. 057103.

Chhabra, R, P 2007, `Bubbles, Drops, and Particles in Non-Newtonian
Fluids', 2\textsuperscript{nd}ed, Taylor \& Francis.

Chou, W,-H \& Faeth, G, M 1998, `Temporal properties of secondary drop
breakup in the bag breakup regime', Int. J. Multiphase Flow, Vol. 24,
pp. 889-912.

Clift, R \& Grace, J, R 1972, `The mechanisms of bubble breakup in
fluidized beds', Ind. Eng. Chem. Fundam., Vol. 27, pp. 2309-2310.

Clift, R, Grace, J, R \& Weber, M, E 1974, `Stability of bubbles in
fluidized beds', Ind. Eng. Chem. Fundam., Vol. 13, pp. 45-51.

Clift, R, Grace, J, R, \& Weber, M, E, 1978, `Bubbles, drops and
particles', Academic Press, Inc, New York.

Cox, R, G 1969, `The deformation of a drop in a general time-dependent
fluid flow', J. Fluid Mech., Vol. 37, pp. 601-623.

Dai, Z, Faeth, G, M 2001, `Temporal properties of secondary drop breakup
in the multimode breakup regime', Int. J. Multiphase Flow, Vol. 27, pp.
217--236.

Davies, J, T 1985, `Drop sizes of emulsions related to turbulent energy
dissipation rates', Chem. Eng. Sci., Vol. 40, pp. 839-842.

De Bruijn, R, A 1989, `Deformation and breakup of drops in simple shear
flows', PhD thesis, Tech. Univ. Eindhoven.

De Bruijn, R, A 1993, `Tipstreaming of drops in simple shear flows',
Chem. Eng. Sci., Vol. 48, pp. 277-284.

Delichatsios, M, A, 1975, `Model for the breakup rate of spherical drops
in an isotropic turbulent flows', Phys. Fluids, Vol. 18, pp. 622-623.

Eggleton, C, D, Tsai, T \& Stebe, K, J 2001, `Tip streaming from a drop
in the presence of surfactants', Phys. Rev. Lett., Vol. 87, Number 4,
pp. 1-4.

Feng J, Q 2010, `A deformable liquid drop falling through quiescent gas
at terminal velocity', J. Fluid Mech., Vol. 658, pp. 438-462.

Frankel, N, A \& Acrivos, A 1970, `The constitutive equation for a
dilute emulsion', J. Fluid Mech., Vol. 44, pp. 65-78.

Galinat, S, Risso, F, Masbernat, O \& Guiraud, P 2007, `Dynamics of drop
breakup in inhomogeneous turbulence at various volume fractions', J.
Fluid Mech., Vol. 578, pp. 85-94.

Gelfand, B, E 1996, `Droplet Breakup Phenomena in Flows with Velocity
Lag', Prog. Energy Comb. Sci., Vol. 22, pp. 201--265.

Goren, S \& Gavis, J 1961, `Transverse wave motion on a thin capillary
jet of a viscoelastic liquid', Phys. Fluids, Vol. 4, pp. 575--579.

Grace, H, P 1971, `Dispersion phenomena in high viscosity immiscible
fluid systems and application of static mixers as dispersion devices in
such systems', Eng. Found., Res. Conf. Mixing, 3\textsuperscript{rd},
Andover, N. H., Republished in 1982, Chem. Eng. Com. Vol. 141, pp.
225-277.

Grace, J, R, Wairegi, T \& Brophy, J 1978, `Break-up of drops in
stagnant media', Can. J. Chem. Eng., Vol. 56, pp. 3-8.

Guildenbecher, D, R 2009, `Secondary Atomization of Electrostatically
Charged Drops', PhD thesis, Purdue University.

Guildenbecher, D, R, López-Rivera \& Sojka, P, E 2009, `Secondary
atomization', Exp Fluids, Vol. 46, pp. 371-402.

Guildenbecher, D, R \& Sojka, P, E 2011, `Experimental investigation of
aerodynamic fragmentation of liquid drops modified by electrostatic
surface charge', Atom. Sprays, Vol. 21, pp. 139--147.

Ha, J.-W \& Yang, S.-M 2000, `Deformation and breakup of Newtonian and
non-Newtonian conducting drops in an electric field', J. Fluid Mech.,
Vol. 405, pp. 131-156.

Haas, F, C, 1964, `Stability of droplets suddenly exposed to a high
velocity gas stream', AIChE J., Vol. 10, pp. 920-924.

Han, J, Tryggvason, G 1999, `Secondary breakup of axisymmetric liquid
drops. I. Acceleration by a constant body force', Physics of Fluids,
Vol. 11(12), pp. 3650-3667.

Han, J, Tryggvason, G 2001, `Secondary breakup of axisymmetric liquid
drops. II. Impulsive acceleration', Phys. Fluids, Vol. 13(6), pp.
1554-1565.

Hanson, A, R, Domich, E, G \& Adams, H, S 1963, `Shock-tube
investigation of the breakup of drops by air blasts', Phys. Fluids, Vol.
6, pp. 1070-1080.

Hesketh, R, P, Russell, T, W, F \& Etchells, A, W 1987, `Bubble size in
horizontal pipelines', AIChE J., Vol. 33, pp. 663-667.

Hesketh, R, P, Etchells, A, W \& Russell, T, W, F 1991a, `Bubble
breakage in pipeline flow', Chem. Eng. Res., Vol. 46, pp. 1-9.

Hesketh, R, P, Etchells, A, W \& Russell, T, W, F 1991b, `Experimental
observations of bubble breakage in turbulent flow', Ind. Eng. Chem.
Res., Vol. 30, pp. 835-841.

Higuera, F, J 2004, `Axisymmetric inviscid interaction of a bubble and a
vortex ring', Phys. Fluids, Vol. 16, No. 4, pp. 1156-1159.

Hinch, E, J \& Acrivos, A 1979, `Steady long slender droplets in
two-dimensional straining motion', J. Fluid Mech., Vol. 91, pp. 401-414.

Hinch, E, J \& Acrivos, A 1980, `Long slender drops in a simple shear
flow', J. Fluid Mech., Vol. 98, pp. 308-328.

Hinze, J, O 1955, `Fundamentals of the hydrodynamic mechanism of
splitting in dispersion processes', AIChE J., Vol. 1, pp. 289-295.

Hsiang, L, P \& Faeth, G, M 1992, `Near-limit deformation and secondary
breakup', Int. J. Multiphase Flow, Vol. 18, pp. 635-652.

Hu, S \& Kinter, R, G 1955, `The fall of single liquid drops through
water', AIChE J., Vol. 1, pp. 43-49.

Jain, M, Prakash, R, S, Tomar, G \& Ravikrishna, R, V 2015, `Secondary breakup of a drop moderate Weber numbers', Proc. R. Soc. A, Vol. 471, pp. 1-25.

Jain, S, S, Tyagi, N, Tomar, G, Prakash, R, S, Ravikrishna, R, V \& Raghunandhan, B, N 2016 'Effect of Density Ratio on the Secondary Breakup of Spherical Drops in a Gas Flow', Presented at 18th Annual Conference of Liquid Atomization \& Spray Systems - Asia, Chennai, India (ILASS - Asia 2016)

Jain, S, S, Tyagi, N, Tomar, G, Prakash, R, S, Ravikrishna, R, V \& Raghunandhan, B, N 2017 'Density ratio and Reynolds number effect in the Secondary breakup of drops at moderate Weber numbers', Manuscript under preparation.

Janssen, J, J, M, Boon, A \& Agterof, W, G, M 1994, `Influence of
dynamic interfacial properties on droplet breakup in simple-shear flow',
AIChE J., Vol. 40, pp. 1929--1939.

Janssen, J, J, M, Boon, A \& Agterof, W, G, M 1997, `Influence of
dynamic interfacial properties on droplet breakup in plane hyperbolic
flow', AIChE J., Vol. 43, pp. 1436--1447.

Kang, I, S \& Leal, L, G 1987, `Numerical solution of axisymmetric,
unsteady free-boundary problems at finite Reynolds number. I. Finite
difference scheme and its application to the deformation of a bubble in
a uniaxial straining flow', Phys. Fluids, Vol. 30, pp. 1929-1940.

Kang, I, S \& Leal, L, G 1989, `Numerical solution of axisymmetric,
unsteady free-boundary problems at finite Reynolds number. II.
Deformation of a bubble in a biaxial straining flow', Phys. Fluids A,
Vol. 1, pp. 644-660.

Kang, I, S \& Leal, L, G 1990, `Bubble dynamics in time-periodic
straining flows', J. Fluid Mech., Vol. 218, pp. 41-69.

Kékesi, T, Amberg, G \& Wittberg, L, P 2014, `Drop deformation and
breakup', Int. J. Multiphase Flow, Vol. 66, pp. 1-10.

Kékesi, T, Amberg, G \& Wittberg, L, P 2016, `Drop deformation and
breakup in flows with shear', Chem. Eng. Sci., Vol. 140, pp. 319-329.

Khakar, D, V \& Ottino, J, M 1987, `Breakup of liquid threads in linear
flows', Int. J. Multiphase Flow, Vol. 13, pp. 147-180.

Klett, J, D 1971, `On the breakup of water drops in air', J. Atmos.
Sci., Vol. 28, pp. 646-647.

Kolmogorov, A, N, 1949, `On the disintegration of drops in a turbulent
flow', Doklady Akad. Nauk., Vol. 66, pp. 825-828.

Komabayasi, M, Gonda, T \& Isono, K 1964, `Life time of water drops
before breaking and size distribution of fragments droplets', J. Met.
Soc. Japan, Vol. 42, pp. 330-340.

Komrakova, A, Shardt, O, Eskin, D, Derksen, J 2015, `Effects of
dispersed phase viscosity on drop deformation and breakup in inertial
shear flow', Chem. Eng. Sci., Vol. 126, pp. 150--159.

Krishna, P, M, Venkateswarlu, D \& Narasimhamurty, G, S, R 1959, `Fall
of liquid drop in water, drag coefficients, peak velocities and maximum
drop sizes', J. Chem. Engng Data, Vol. 4, pp. 340-343.

Kubie, J. \& Gardner, G, C 1977, `Drop sizes and drop dispersion in
straight horizontal tubes and in helical coils', Chem. Eng. Sci., Vol.
32, pp. 195-202.

Lamb, H, 1932, `Hydrodynamics', Cambridge University Press.

Lane, W, R 1951, `Shatter of drops in streams of air', Ind. Eng. Chem.,
Vol. 43, pp. 1312-1317.

Levich, V, G 1962, `Physico-chemical hydrodynamics', Prentice-Hall,
Englewood Cliffs. N. J.

Lewis, D, A \& Davidson, J, F 1982, `Bubble splitting in shear flow',
Trans. IChemE, Vol. 60, pp. 283-291.

Li, J, Renardy, Y, Renardy,M, 2000, `Numerical simulation of breakup of
a viscous drop in simple shear flow through a volume-of-fluid method',
Phys. Fluids, Vol. 12 (2), pp. 269-282.

Liao, Y \& Lucas, D, 2009, `A literature review of theoretical models
for drop and bubble breakup in turbulent dispersions', Chem. Eng. Sci.,
Vol. 64, pp. 3389-3406.

Liao, Y \& Lucas, D, 2010, `A literature review on mechanisms and models
for the coalescence process of fluid particles', Chem. Eng. Sci., Vol.
65, pp. 2851-2864.

Liao, Y, Lucas, D, Krepper, E \& Schmidke, M 2011, `Development of a
generalized coalescence and breakup closure for the inhomogeneous MUSIG
model', Nuclear
Engineering and Design, Vol. 241, pp. 1024-1033.

Liu, Z, Reitz, R, D 1997, `An analysis of the distortion and breakup
mechanisms of high speed liquid drops', Int. J. Multiphase Flow, Vol.
23(4), pp. 631-650.

Martinez-Bazan, C, Montanes, J, J \& Lasheras, J, C 1999, `On the
breakup of an air bubble injected into a fully developed turbulent flow.
Part 1. Breakup frequency', J. Fluid Mech., Vol. 401, pp. 157-182.

Mikami, T, Cox, R \& Mason, S, G 1975, `Breakup of extending liquid
threads', Int. J. Multiphase Flow, Vol. 2, pp. 113-138.

Misksis, M, A 1981, `A bubble in an axially symmetric shear flow', Phys.
Fluids, Vol. 24, pp. 1229-1231.

Ohta, M, Imura, T, Yoshida, Y \& Sussman, M 2005, `A computational study
of the effect of initial bubble conditions on the motion of a gas bubble
rising in viscous liquids', Int. J. Multiphase Flow, Vol. 31, pp.
223-237.

Padrino, J, C \& Joseph, D, D 2011, `Viscous irrotational analysis of
the deformation and break-up time of a bubble or drop in uniaxial
straining flow', J. Fluid Mech., Vol. 688, pp. 390-421.

Paul, H, I \& Sleicher, C, A 1965, `The maximum stable drop size in
turbulent flow: effect of pipe diameter', Chem. Eng. Sci., Vol. 20, pp.
57-59.

Pilsh, M \& Erdman, C, A 1987, `Use of breakup time data velocity
history data to predict the maximum size of stable fragments for
acceleration-induced breakup of a liquid drop', Int. J. Multiphase Flow,
Vol. 13, pp. 741-757.

Pimbley, W, T \& Lee, H, C 1977, `Satellite droplet formation in a
liquid jet', IBM J. Res. Dev., Vol. 21, pp. 21-30.

Plateau, J 1873, `Statique experimentale et theorique des liquids soumis
aux seules forces moleculaires', Cited by Rayleigh, L 1945, `Theory of
Sound', Vol. 2, pp. 363. New York: Dover.

Ponstein, J 1959, `Instability of rotating cylindrical jets', Appl. Sci.
Res., Vol. 8, pp. 425--456.

Prakash, R, S, Jain, S, S, Tomar, G, Ravikrishna, R, V, Raghunandhan, B, N 2016 'Computational study of liquid jet breakup in swirling cross flow', Presented at 18th Annual Conference of Liquid Atomization \& Spray Systems - Asia, Chennai, India (ILASS - Asia 2016)

Qian, D, McLaughlin, J, B, Sankaranarayanan, K, Sundaresan, S \&
Kontomaris, K 2006, `Simulation of bubble breakup dynamics in
homogeneous turbulence', Chem. Eng. Comm., Vol. 193, pp. 1038-1063.

Rallison, J, M \& Acrivos, A 1978, `A numerical study of the deformation
and burst of a viscous drop in an extensional flow', J. Fluid Mech.,
Vol. 89, pp. 191-200.

Rallison, J, M 1981, `A numerical study of the deformation and burst of
a viscous drop in general shear flows', J. Fluid Mech., Vol. 109, pp.
465-482.

Rallison, J, M 1984, `The deformation of small viscous drops and bubbles
in shear flows', Ann. Rev. Fluid Mech., Vol. 16, pp. 45-66.

Ravelet, F, Colin, C and Risso, F 2011, `On the dynamics and breakup of
a bubble rising in a turbulent flow', Physics of Fluids, Vol. 23, pp.
1-12.

Rayleigh, L 1878, `On the instability of jets', Proc. London Math. Soc.,
pp. 4-13.

Rayleigh, L 1892, `On the instability of cylindrical fluid surfaces',
Phil. Mag, Vol. 34, pp. 177-180.

Renardy, Y, Cristini, V, 2001a, `Effect of inertia on drop breakup under
shear', Phys. Fluids, Vol. 13(1), pp. 7-13.

Renardy, Y, Cristini, V, 2001b, `Scalings for fragments produced from
drop breakup in shear flow with inertia', Phys. Fluids, Vol. 13(8), pp.
2161-2164.

Renardy, Y, Cristini, V, Li, J, 2002, `Drop fragment distributions under
shear with inertia', Int. J. Multiphase Flow, Vol. 28, pp. 1125--1147.

Renardy, Y, 2008, `Effect of startup conditions on drop breakup under
shear with inertia', Int. J. Multiphase Flow, Vol. 34, pp. 1185--1189.

Revuelta, A, Rodriguez-Rodriguez, J \& Martinez-Bazan, C 2006, `Bubble
break-up in a straining flow at finite Reynolds numbers', J. Fluid
Mech., Vol. 551, pp. 175-184.

Revuelta, A, Rodriguez-Rodriguez, J \& Martinez-Bazan, C 2008, `On the
breakup of bubbles at high Reynolds numbers and subcritical Weber
number', Eur. J. Mech. B/Fluids, Vol. 27, pp. 591-608.

Revuelta, A 2010, `On the interaction of a bubble and a vortex ring at
high Reynolds numbers', Eur. J. Mech. B/Fluids, Vol. 29, pp. 119-126.

Rimbert, N \& Castanet, G 2011, `Crossover between Rayleigh-Taylor
instability and turbulent cascading atomization mechanism in the
bag-breakup regime', Phys. Rev. E, Vol. 84.

Risso, F \& Fabre, J 1998, `Oscillations and breakup of a bubble
immersed in a turbulent field', J. Fluid Mech., Vol. 372, pp. 323-355.

Risso, F 2000, `The mechanisms of deformation and breakup of drops and
bubbles', Multiphase Sci. Technol., Vol. 12, pp. 1-50.

Rodriguez-Rodriguez, J, Gordillo, J, M \& Martinez-Bazan, C 2006,
`Breakup time and morphology of drops and bubbles in a
high-Reynolds-number flow', J. Fluid Mech., Vol. 548, pp. 69-86.

Rumscheidt, F, D \& Mason, S, G 1961, `Particle motions in sheared
suspensions. XII. Deformation and burst of fluid drops in shear and
hyperbolic flows', J. Colloid Interf. Sci., Vol. 16, pp. 238-261.

Rumscheidt, F, D \& Mason, S, G 1962, `Break-up of stationary liquid
threads', J. Colloid Science, Vol. 17, pp. 260-269.

Rutland, D, F \& Jameson, G, J 1970, `Theoretical prediction of the
sizes of drops formed in the breakup of capillary jets', Chem. Eng.
Sci., Vol. 25, pp. 1689--1698.

Ryan, R, T 1978, `The possible modification of convective systems by the
use of surfactants', J. Appl. Met., Vol. 15, pp. 3-8.

Ryskin, G \& Leal, L, G 1984, `Numerical solution of free-boundary
problems in fluid mechanics. Part 3. Bubble deformation in an
axisymmetric straining flow', J. Fluid Mech., Vol. 148, pp. 37-43.

Sathyagal, A, N, Ramkrishna, D \& Narsimhan, G 1996, `Droplet breakage
in stirred dispersions. Breakage functions from experimental drop-size
distributions', Chem. Eng. Sci., Vol. 51, pp. 1377-1391.

Sevik, M \& Park, S, H 1973, `The splitting of drops and bubbles by
turbulent fluid flow', J. Fluids Engng., pp. 53-60.

Shinnar, R 1961, `On the behavior of liquid dispersions in mixing
vessels', J. Fluid Mech., Vol. 10, pp. 259-275.

Sleicher, C, A 1962, `Maximum stable drop size in turbulent flow', AIChE
J., Vol. 8, pp. 471-477.

Solsvik, J, Tangen, S \& Jakobsen, H A 2013 `On the constitutive
equations for fluid particle breakage', Rev. Chem. Eng., Vol. 29, pp.
241-356.

Spangler, C, A, Hibling, J, H \& Heister, S, D, `Nonlinear modeling of
jet atomization in the wind-induced regime', Phys. Fluids, Vol. 7(5),
pp. 964-971.

Stone, H, A, Bentley, B, J \& Leal, L, G 1986, `An experimental study of
transient effects in the breakup of viscous drops', J. Fluid Mech., Vol.
173, pp. 131-158.

Stone, H, A \& Leal, L, G 1989, `Relaxation and breakup of an initially
extended drop in an otherwise quiescent fluid', J. Fluid Mech., Vol.
198, pp. 399-427.

Stone, H. A. \& Leal, L. G 1990, `The effects of surfactants on drop
deformation and breakup', J. Fluid Mech., Vol. 220, pp. 161-186.

Stone, H, A 1994, `Dynamics of drop deformation and breakup in viscous
fluids', Annu. Rev. Fluid Mech., Vol. 26, pp. 65-102.

Taylor, G, I 1932, `The viscosity of a fluid containing small drops of
another fluid', Proc. R. Soc. A, Vol. 138, pp. 41-48.

Taylor, G, I 1934, `The formation of emulsions in definable fields of
flow', Proc. R. Soc. A, Vol. 146, pp. 501-523.

Taylor, G, I 1964, `Conical free surfaces and fluid interfaces', Proc.
Int. Cong. Of Appl. Mech., 11\textsuperscript{th}, Munich, pp. 790-796.

Tjahjadi, M, Stone, H, A \& Ottino, J, M 1992, `Satellite and
subsatellite formation in capillary breakup', J. Fluid Mech., Vol. 243,
pp. 297-317.

Tomotika, S 1935, `On the stability of a cylindrical thread of a viscous
liquid surrounded by another viscous fluid', Proc. R. Soc. Lond. Ser. A,
Vol. 150, pp. 322-337.

Tomotika, S 1936, `Breaking up of a drop of viscous liquid immersed in
another viscous fluid which is extending at a uniform rate', Proc. R.
Soc. Lond. Ser. A, Vol. 153, pp. 302-318.

Torza, S, Cox, R, G \& Mason, S, G 1972, `Particle motions in sheared
suspension. XXVII. Transient and steady deformation and burst of liquid
drops', J. Colloid Int. Sci., Vol. 38, pp. 395-411.

Tripathi, M, K, Sahu, K, C \& Govindarajan, R 2015, `Dynamics of an
initially spherical bubble rising in quiescent liquid', Nature
Communications, Vol. 6, pp. 1-9.

Wallis, G, B 1969, `One-dimensional two-phase flow', McGraw Hill.

Walter, J, F \& Blanch, H, W 1986, `Bubble break-up in gas-liquid
bioreactors: break-up in turbulent flows', Chem. Eng. J., Vol. 32, pp.
7-17.

Weber, C 1931, `On the breakdown of a fluid jet, Zum Zerfall eines
Flussigkeitsstrahles', Z. Angew. Math. und Mech., Vol. 11, pp. 136--154.

Xiao, F, Dianat, M, \& McGuirk, J, J 2014, `Large eddy simulation of
single droplet and liquid jet primary breakup using a coupled level set/
volume of fluid method', Atomization Spray, Vol. 24, pp. 281-302.

Youngren, G, K \& Acrivos, A 1976, `On the shape of a gas bubble in a
viscous extensional flow', J. Fluid Mech., Vol. 76, pp. 433-442.

Yu, K , L 1974, PhD thesis, University of Houston.